\documentclass[]{aastex631}

\usepackage{soul}
\usepackage{xcolor}
\usepackage[T1]{fontenc}
\usepackage{CJK}
\usepackage{multirow}
\begin{document}

\title{Monte Carlo simulation of the Compton scattering and disk reflection of a cylinder with hot electrons moving away from a black hole}

\correspondingauthor{Yuan You}
\email{youyuan@ihep.ac.cn}

\correspondingauthor{Shuang-Nan Zhang}
\email{zhangsn@ihep.ac.cn}

\author[0009-0002-3003-2559]{Wei Meng}
\affiliation{Key Laboratory of Particle Astrophysics, Institute of High Energy Physics, Chinese Academy of Sciences, Beijing 100049, People's Republic of China}
\affiliation{University of Chinese Academy of Sciences, Chinese Academy of Sciences, Beijing 100049, People's Republic of China}

\author[0000-0002-0352-8148]{Yuan You}
\affiliation{Key Laboratory of Particle Astrophysics, Institute of High Energy Physics, Chinese Academy of Sciences, Beijing 100049, People's Republic of China}

\author[0000-0001-5586-1017]{Shuang-Nan Zhang}
\affiliation{Key Laboratory of Particle Astrophysics, Institute of High Energy Physics, Chinese Academy of Sciences, Beijing 100049, People's Republic of China}
\affiliation{University of Chinese Academy of Sciences, Chinese Academy of Sciences, Beijing 100049, People's Republic of China}

\author[0009-0002-5762-7384]{Jia-Ying Cao}
\affiliation{Key Laboratory of Particle Astrophysics, Institute of High Energy Physics, Chinese Academy of Sciences, Beijing 100049, People's Republic of China}
\affiliation{University of Chinese Academy of Sciences, Chinese Academy of Sciences, Beijing 100049, People's Republic of China}

\begin{abstract}

For accreting black holes (BHs), the lamp-post scenario is a simple and popular model: a hot and point-like corona is located above the black hole, irradiating the accretion disk with hard X-ray radiation, which is believed to be generated by inverse Compton scattering in the corona. Although the lamp-post model successfully explains the disk reflection component, it fails to address the origin of seed photons and the geometry of the hot corona, because this model simplistically treats the corona as a point-like source generating a cutoff powerlaw spectrum.
% Despite that the lamp-post scenario can successfully interpret the disk reflection component, it avoids the source of seed photons and the geometry of the hot corona, since the hot corona is simplified as a point-like source producing a cut-off powerlaw spectrum.
In this paper, we make simulations on a possible physical realization of the lamp-post scenario: the shape of the scattering zone is set to be a cylinder, corresponding to the jet base near a BH. The source of seed photons in this system is assumed to be the multicolor blackbody radiation of the accretion disk. In our simulations, the Compton scattering process is simulated with a custom Monte Carlo program based on the Geant4 package and the disk reflection process is simulated with the \textbf{xilconv} model in XSPEC. Our simulation results have confirmed that the relativistic motion of the jet can weaken or even completely suppress the reflection of the accretion disk, and simultaneously, the Comptonization of disk photons in the jet can still make a major contribution to the observed X-ray spectrum in high energy range.
We discuss the implications of our simulation results, in light of the recent observations of a very weak reflection component in the presence of a strong hard X-ray radiation from the outburst of Swift J1727.8-1613.

\end{abstract}

\section{Introduction} \label{sec:intro}

Black holes (BHs), as the most extreme celestial bodies, have long attracted extensive attention. Usually, BHs in X-Ray Binaries (XRBs) are surrounded by their accretion disks, which can generate thermal radiation because of the hot temperature of their inner region \citep{Shakura}. Another structure possibly exists near the BH is the hot corona, i.e., a highly ionized plasma region \citep{Galeev,Haardt,Poutanen}. The hot electrons in the corona can up-scatter some low-energy photons and generate the so-called non-thermal or Compton component in the observed spectra \citep{Sunyaev1979}. In some cases, these Comptonized photons may illuminate the accretion disk and finally be reflected towards the observer \citep{Lightman1988,Magdziarz1995,Ross,Garc2010}. It is known that the disk is filled with semi ionized hydrogen, helium, and other metallic elements; therefore, the reflection component will have various emission lines and a Compton hump. The disk component, the Compton component, and the reflection component together compose a typical accreting BH spectrum \citep{Bambi}.

There are many models about the geometry of the corona near an accreting BH. For instance, a slab-like corona that sandwiches the accretion disk, or a hot inner accretion flow \citep{Bambi}. Among these models, the lamp-post scenario is a popular model which describes the irradiation of the corona to the disk. In this scenario, the hot corona is located above the BH, illuminating the disk with hard X-ray radiation \citep{Matt1991,Martocchia1996,Miniutti2004,Ursini2020}.
The lamp-post scenario can interpret the properties of observed disk reflections in general.
With the hot corona simplified as a point-like hard X-ray source, the typical reflection characteristics such as emission lines, Compton hump, and angular dependence can be well simulated (e.g. the \textbf{xillver} model, \citealt{Garc2013}); and in recent reflection models, the relativistic effects can also be included (e.g. the \textbf{relxill} model, \citealt{Garc2014}). These works make the reflection calculations accurate enough to estimate the inner disk radius and further the spin of the BH.

Despite the huge progress based on the lamp-post model, the physical processes inside the lamp-post corona cannot be further studied due to the assumed point-like geometry. The lamp-post scenario also leads  inevitably to a serious tension between the reflection flux spectral fitting and the time-lag spectral fitting \citep{Zoghbi2020,Mastroserio2020,Wang2021}. Extended coronal geometries are typically modeled as spherical or slab-like configurations, and correlation analysis of the disc reflection spectrum for extended coronal geometries was performed by \cite{Wilkins2012,Wilkins2017} and \cite{Zhang2024}. Nevertheless, it is commonly considered that the lamp-post corona corresponds to the base of the jet near the BH \citep{Markoff}. This is supported by the timing and spectral analyses of MAXI J1820+070, in which the high energy quasi-periodic oscillations above 200 keV are modelled as the precession of an X-ray jet \citep{Ma2021} and decrease of the reflection component is believed to be caused by the increase of the bulk velocity of the jet \citep{You}. \citet{Lucchini2023} also reported that a vertically extended corona, mimicking a jet base, can capture both timing and spectral properties of MAXI J1820+070 near its hard-to-soft state transition, while a lamp-post corona cannot. The observations of Swift J1727.8-1613 exhibit an extra hard component other than the reflection component, which is also consistent with a jet corona model \citep{Peng}. As for the possible sources of seed photons to the hard component, there are generally two possibilities: the external source or internal source of seed photons. The external source is normally considered as the thermal radiation of the disk, while the internal source is assumed to be the Synchrotron radiation inside the corona itself, the so-called Synchrotron Self-Compton (SSC) process which was invoked to explain the weak reflection component observed in Swift J1727.8-1613 \citep{Peng}. It is argued previously that if the hard X-ray radiation from a jet-like corona can be weakly reflected by the disk due to special relativistic effects, then it is also hard for a jet-like corona to intercept the disk photons due to the same special relativistic effects \citep{Beloborodov1999}. It seems that internal source of seed photons is more acceptable, yet further in-depth studies are needed to clarify this issue.

In this paper, we propose a model as a physical realization of the lamp-post scenario and test it with Monte Carlo (MC) simulations, in order to explore how a weak disk reflection and a strong hard X-ray radiation can coexist in a BH-jet system, and whether an external accretion disk as the only source is able to offer sufficient seed photons for the Comptonization process. The MC simulation program is based on Geant4, an open-source Monte Carlo toolkit developed by the European Council for Nuclear Research (CERN). As the first of a series of works, we make a simple model, in which the seed photons are emitted from the accretion disk, and are scattered by a cylindrical corona at the lamp-post position, mimicking the jet base near the BH. The subsequent disk reflection of the hard X-ray radiation of the corona is taken into consideration. The Compton scattering process is calculated by a Geant4 based simulating program \citep{Hou2022}, and the reflection process is estimated by an XSPEC model. In Section \ref{sec:MONTE CARLO SIMULATION}, we introduce our simulation in detail, including the geometry settings, the Compton scattering MC program and the reflection model. Section \ref{sec:results} shows the simulation results. The influences of the jet velocity and other factors on the Compton and reflection components in the simulated spectra are discussed in Section \ref{sec:disc}. Finally, a summary is made in Section \ref{sec:Summary}.

\section{simulation settings} \label{sec:MONTE CARLO SIMULATION}

\subsection{Compton MC program based on Geant4}

Geant4 is an open-source Monte Carlo toolkit written in C++, developed by CERN  \citep{Geant4_1,Geant4_2,Geant4_3}. Usually, it is utilized in simulating particle physics experiments or developing related instruments such as accelerators, detectors, etc. We choose to use Geant4 to build our custom astrophysical MC program for two reasons. First, Geant4 is an advanced and mature software, which contains complete MC simulating processes and its geometry module can help us build the mass model of an object with whatever shape or material, like building blocks. Second, Geant4 is open-source and consists of standard modules, which is quite convenient for us to add custom modifications.

\begin{figure}[htbp!]
    \centering
    \includegraphics[width=0.75\linewidth]{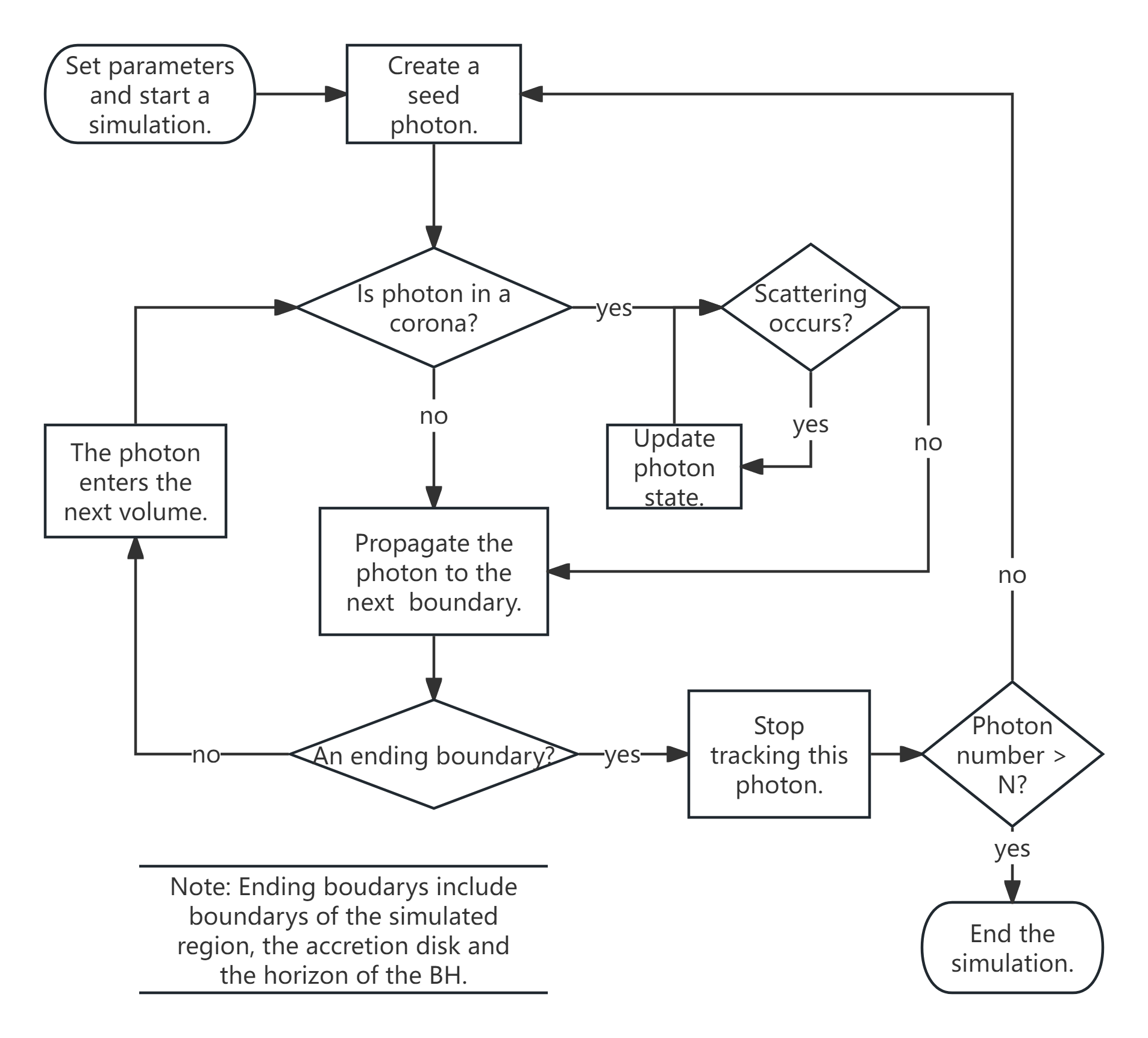}
    \caption{A flow chart of the Comptonization MC program.}
    \label{fig:workflow}
\end{figure}

To compute the Comptonization process of the disk radiation, a custom inverse-Compton scattering reaction class is derived from the template reaction class of Geant4. This Geant4 custom reaction class considers an idealized situation where traced photons undergo the Compton scattering interaction with free electrons exclusively, so that there is no photon absorption and the photon number is strictly conserved. In this custom reaction class, we transform to the electron rest frame (by Lorentz transformation) and use Klein-Nishina formula to compute the inverse Compton process; this Compton scattering Monte Carlo algorithm has been validated in a previous study \citep{Hou2022}, in which reference frame transformation, cross section sampling, corona electron sampling and other calculation processes are described in details. The materials and shapes of the scattering corona can be defined by following the normal usage of Geant4, similar with building a mass model of a detector in the general use of Geant4. As a result, this MC program is capable of simulating the Comptonization process in electron corona of arbitrary shape and bulk motion \footnote{GitLab: \url{https://code.ihep.ac.cn/youyuan/freecomp}}. The general workflow of this Geant4 MC program is shown in Figure \ref{fig:workflow}.
%After sampling seed photons, the processing flow for each seed photon is as follows. First, the code determines whether the seed photon will undergo inverse Compton scattering. If scattering does not occur, the final state of the photon will be recorded; these photons correspond to the observed disk component. For photons to be scattered, the code calculates their scattering position, the exit energy and the exit direction, then updates the state of the photon and checks again for further scattering. This checking and scattering cycle keeps going on until no further scattering is needed. For photons that are no longer scattered, the code records their final states and checks whether they will return to the accretion disk. Photons scattered and transported to infinity correspond to the Compton component in the observed radiation. Those photons scattered and returning to the disk will be used as the input of an XSPEC model to calculate the reflected radiation.

\subsection{Model Geometry}
\label{Model Geometry}

\begin{figure}[htbp!]
    \centering
    \includegraphics[width=0.75\linewidth]{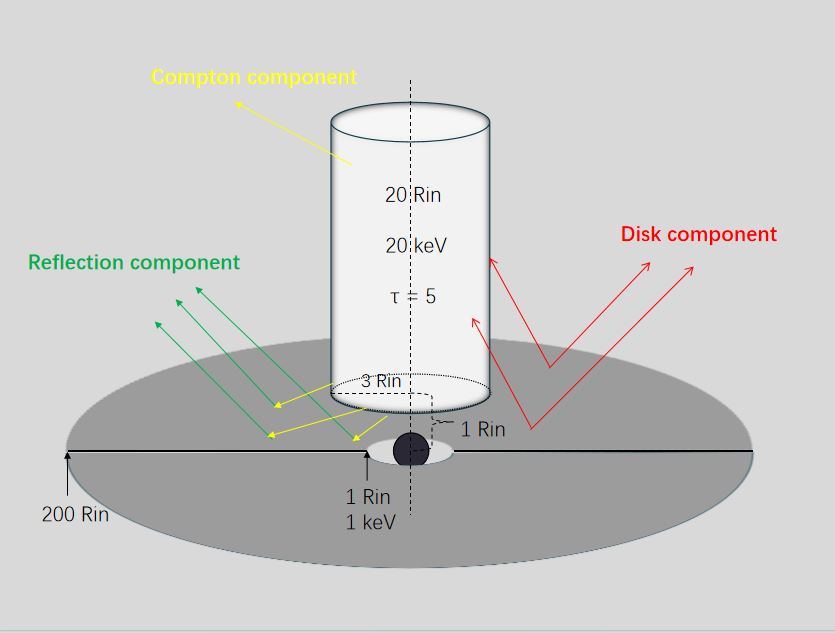}
    \caption{A diagram of the simulated geometry, with some parameters marked on.}
    \label{fig:geometry}
\end{figure}

In previous accreting BH studies, especially those studies on the reflection of the accretion disk, the lamp-post corona is simplified as a point-like light source above the disk on the rotation axis of the disk \citep{Matt1991,Martocchia1996,Miniutti2004,Ursini2020}. However, when considering the scattering process, the shape of the corona becomes important. In this work, as shown in Figure \ref{fig:geometry}, we set the shape of the corona as a cylinder whose axis coincides with the rotation axis of the accretion disk, mimicking the jet base near the BH; this could be a physical realization of the lamp-post scenario. The size of the jet base can vary from 10 to $100R_{\rm{g}}$ and the ratio between the height and radius of the jet base can vary from 0.1 to 100 \citep{Maitra2009}; the temperature range of the corona is approximately $10^8$ to $10^9\ \rm{K}$\citep{Hubeny2001}; here we set the height and radius of the jet base, i.e., the cylindrical corona, to a moderate value: the jet base is located at a vertical distance $h=1R_{\rm{in}}$ from the black hole center (the origin), a radial extent $R_{\rm{corona}}=3R_{\rm{in}}$, and a height $H_{\rm{corona}}=20R_{\rm{in}}$, with a temperature of $20\ \rm{keV}$ and an optical depth $\tau$ which is defined along the path from the center to the edge of the corona (the coronal radius), with a value of 5. Meanwhile, the source of seed photons is another key point when considering scattering processes. Usually, the multicolor blackbody radiation from the accretion disk and the synchrotron radiation of electrons within the corona are two most possible sources of the seed photons.

In this work, we assume that all seed photons originate from the multicolor blackbody radiation emitted by a flat and thin accretion disk, with the temperature at the inner disk radius $1R_{\rm{in}}$ set to 1 $\rm{keV}$, following the relationship $T \propto R^{-3/4}$, and the accretion disk is a geometrically thin ring with an inner disk radius of $1R_{\rm{in}}$ and an outer disk radius of $ 200R_{\rm{in}}$, which is large enough for the accuracy of the simulation result, see details in appendix \ref{appendix Rout}.
Additionally, since our geometric model is constructed in flat spacetime, we use the event horizon to describe the absorption of photons by the black hole, without accounting for the final state of photons that move into the event horizon region. The central celestial body is a BH with a horizon radius of $R_{\rm{in}}/3$. In our simulations, the $R_{\rm{in}}$ of the accretion disk is set as 3 times the horizon radius of the BH, which has the same ratio with a Schwarzschild spacetime, and the horizon is treated as an absolute absorption sphere surface in a Minkowski spacetime for photons. Since that our simulation code is not able to calculate GR effects, the spin parameter is meaningless here, only the geometric structure (ratio between $R_{\rm{in}}$ and $R_{\rm{s}}$) matters. In summary, our geometric model is constructed in flat spacetime and consists of three main components: the accretion disk, the jet/corona, and the BH, as shown in Figure \ref{fig:geometry}.

\begin{figure}[htbp!]
    \centering
    \includegraphics[width=0.85\linewidth]{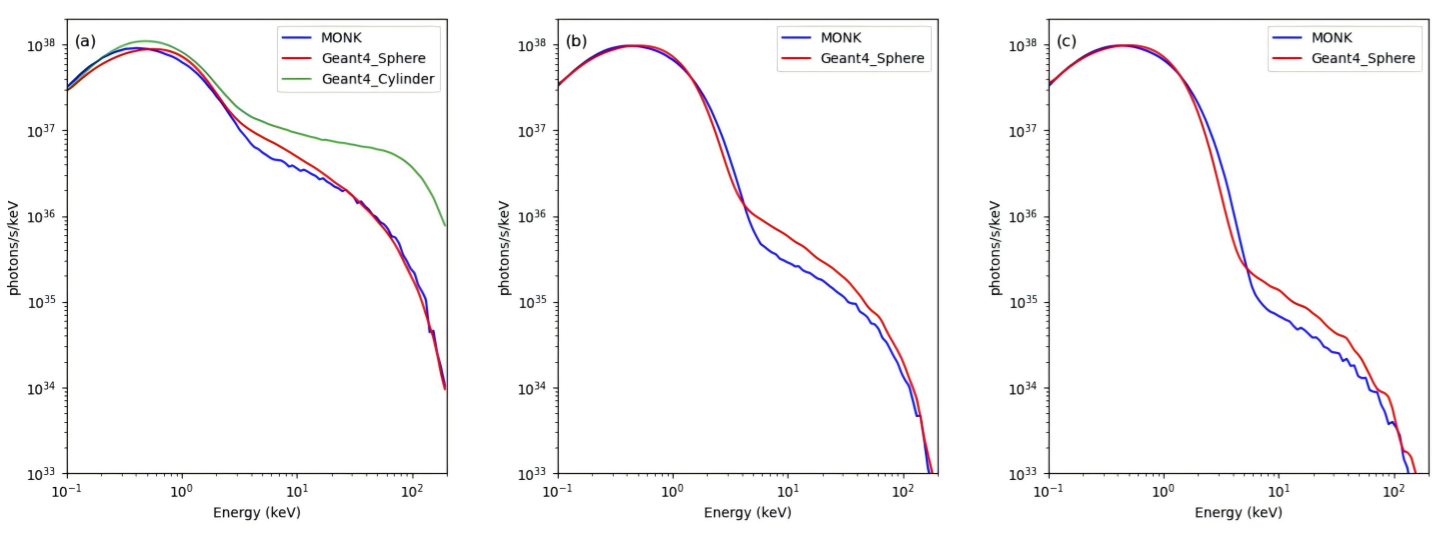}
    \caption{Comparisons between MONK and our Geant4 simulated spectra. The MONK simulations are performed under Kerr spacetime, with the geometry that a spherical corona locates above the accretion disk on the spin axis of the BH, the corona temperature is 20 keV, its height is 24 $M$ ($G=c=1$), its radial optical depth is 5, its bulk-motion velocity away from the disk plane is $0.4 c$, the inclination angle is 65°, and the inner radius of the disk is fixed at $6 M$. The BH parameters are $a=0.9$, $\dot{m}=0.1$, $M_{\odot}=10$. The corona radii are 18 $M$, 6 $M$, 3 $M$ in panels (a), (b) and (c) respectively. For Geant4 simulations, the same geometry and parameter values with MONK are shared to make comparisons, except that the spacetime in Geant4 simulations are flat. Additionally, in panel (a), spectrum of a cylindrical geometry, with its geometry described in Section 2.2, is also plotted. For all simulations in this figure, only disk components and Compton components are calculated, but the reflections are not.}
    \label{geant4_monk}
\end{figure}

It should be noted that in our Geant4 code, the simulation is performed under a flat spacetime, in other words the general relativistic effects caused by strong gravitational fields are neglected. In order to test whether GR effects are significant to our simulation, results of our flat spacetime Geant4 code and those of MONK \citep{Zhang2019}, a Kerr spacetime Compton scattering Monte Carlo code, are compared together, as shown in Figure \ref{geant4_monk}. Since MONK cannot build the corona geometry arbitrarily, we pick a geometry offered by MONK and similar to our cylindrical geometry, that a spherical corona locates above the accretion disk on the spin axis of the BH. Meanwhile, we build the same spherical geometry in our Geant4 code to make comparisons. In MONK, the black hole at the center is set to have spin of 0.9 and mass of 10 $M_\odot$. Such a high spin value is matched with the high-spin favor of the current stellar mass BH spin measurement \citep{BHspin2021}, and it can clearly show to what degree the spectral difference could be, between a flat spacetime simulation and a Kerr one. The inner and outer radii of the disk are set to 6 $M(c=G=1)$ and 180 $M$. From panel (a) to panel (c) in Figure \ref{geant4_monk}, the radius of the corona decreases gradually. When the scale of the corona is much larger than $R_{\rm{in}}$ of the disk, which is the case of panel (a), the difference of spectral shape at high energy range between MONK and our Geant4 code is quite small, while the discrepancy at low energy range can be explained by gravitational redshift. However, when the corona scale became smaller, the difference between MONK and our Geant4 result becomes more and more obvious from panel (b) to (c), because the intensity of the Compton component decreases with the size of the corona, and for a small corona, even a small difference on the ratio of disk radiation entering the corona can cause large relative difference on the intensity of the Compton component.
Thus, it can be concluded that for the case of our large-corona geometry settings, the influence of GR effects on the Compton component of simulated spectrum is not significant. In addition, due to the differences between the MONK spherical geometry and the cylindrical geometry used in this work, we also compared the differences between these two geometries. In panel (a) of Figure \ref{geant4_monk}, the spectrum of a cylindrical corona, whose shape is described in Section 2.2 and with other settings the same with the spherical corona described above, is also plotted, indicating that a cylindrical corona can generate much stronger Compton component than a spherical corona.

\subsection{Seed photons from the Accretion Disk} \label{program}

In our simulations, it is assumed that the seed photons are emitted from the accretion disk and the disk conforms to the multicolor blackbody model. The multicolor blackbody spectrum of the accretion disk is the accumulation of blackbody spectra at different disk radii \citep{Mitsuda}. The temperature of the accretion disk follows the relation $T \propto R^{-3/4}$ and the blackbody radiation formula is given by:
\begin{equation}
B_v (T)=\frac{2hv^3}{c^2}\frac{1}{e^{\frac{hv}{kT}}-1},
\end{equation}
so that the spectra of seed photons at different radii are decided. Additionally, we need to know the emitting density $N$ of the seed photons in the accretion disk, so that the position of a seed photon to be sampled can be decided. For a blackbody radiation, the photon generating rate $n$ is:
\begin{eqnarray}
n = & \int d\Omega\frac{1}{c^{3}}\int\nu^{2}d\nu\frac{1}{e^{\frac{h\nu}{kT}}-1}\nonumber\\
= & \frac{4\pi}{c^{3}}\left(\frac{kT}{h}\right)^{3}\int_{0}^{\infty}\frac{x^{2}dx}{e^{x}-1}\nonumber\\
= & \frac{\zeta(3)}{\pi^{2}c^{3}}\left(\frac{kT}{\hbar}\right)^{3},
\end{eqnarray}
which is proportional to the cube of the temperature: $n \propto T^{3}$. The emitting area emitting at different radii in the accretion disk is proportional to the radius $\delta S \propto R$. Based on these relationships, we obtain the relation between $N$ and $R$ as:
\begin{equation}
    N = n \cdot \delta S \propto T^{3} \cdot R \propto R^{-\frac{5}{4}},
\end{equation}
which is used to sample the position of the seed photons. For the seed photon energy, we set the value of the color-temperature correction factor to 1 and the photon sampling aligns with the model \textbf{diskbb} \citep{Mitsuda,Makishima1986}. Finally, according to the properties of the blackbody radiation, the angular distribution of the emitted seed photons is proportional to $\cos\theta$, where $\theta$ is the angle between the seed photon direction and the normal vector of the disk plane. Based on the above relations, we use the MC method to sample the initial position, energy, and velocity direction of each seed photon.

% It is known that for the accretion disk, the larger the radius, the more photons are emitted. Therefore, simulation with larger $R_{\rm{out}}$ can generate more accurate spectral result theoretically. $R_{\rm{out}}$ is set as $200R_{\rm{in}}$ in our simulation, which needs to be tested whether this size of $R_{\rm{out}}$ is enough for the accuracy of spectrum at the concerned energy range of $0.1-200$ keV. We compared the energy spectra of the disk radiation and the Compton scattering with outer disk radii of $200R_{\rm{in}}$ and $500R_{\rm{in}}$, as presented in Figure \ref{fig:difr}. It is shown that there is no obvious difference between spectra of $200R_{\rm{in}}$ and $500R_{\rm{in}}$, both for the disk radiation and for the Compton scattering. Therefore, an outer disk radius $200R_{\rm{in}}$ is acceptable for our simulations.

% \begin{figure}[htbp!]
%     \centering
%     \includegraphics[width=0.75\linewidth]{difr.png}
%     \caption{Comparisons of the energy spectra of disk and Compton scattering between disk outer radii of $200R_{\rm{in}}$ and $500R_{\rm{in}}$, in energy range $0.1-200$ keV. The left panel shows the energy spectra of direct disk radiation without Compton scattering with different outer disk radii. The right panel shows the energy spectra of pure Compton radiations with different outer disk radii.}
%     \label{fig:difr}
% \end{figure}

\subsection{the Reflection Model}

\begin{figure}[htbp!]
    \centering
    \includegraphics[width=0.75\linewidth]{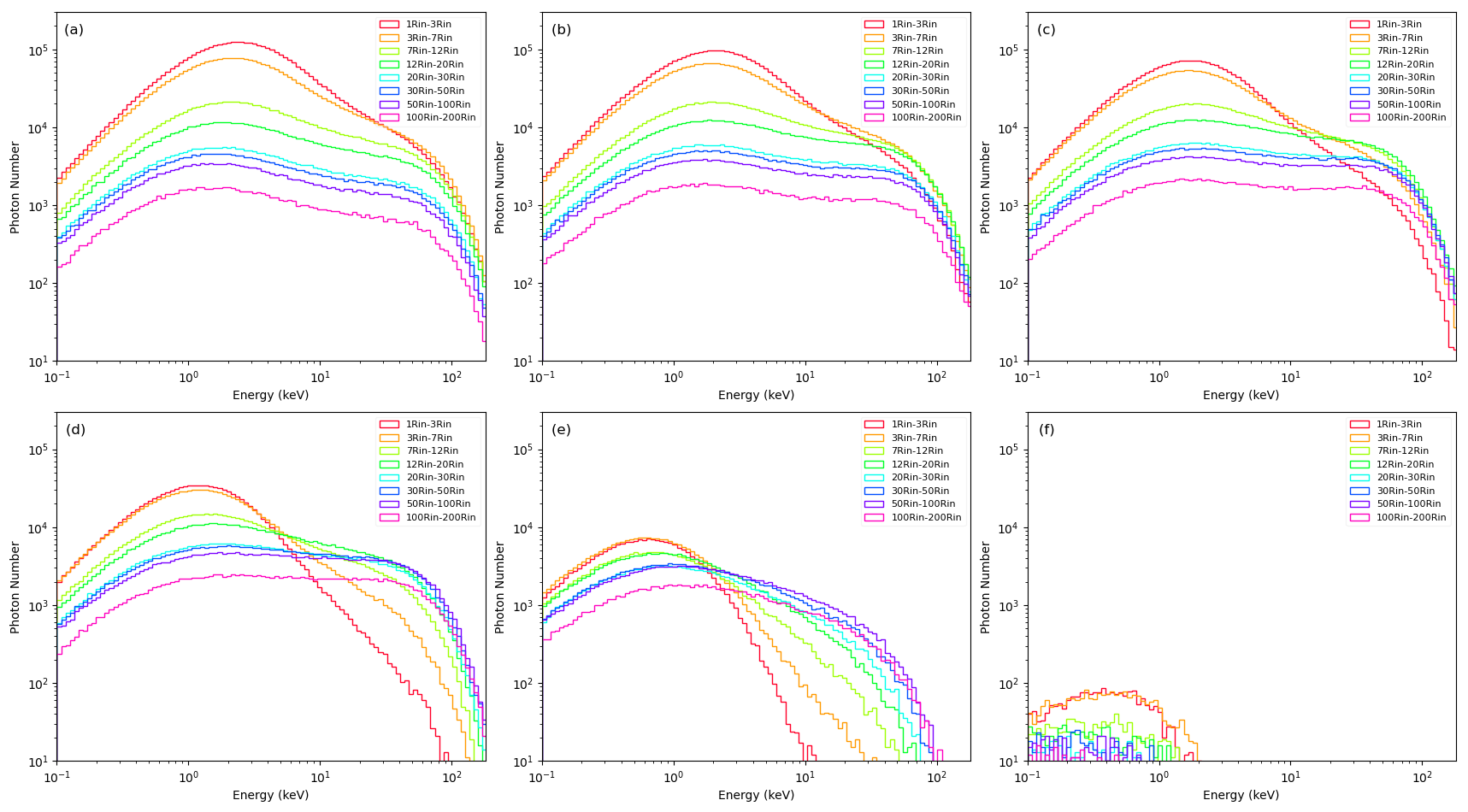}
    \caption{The spectra of return-to-disk photons in the eight radial bins ($1R_{\rm{in}}$-$3R_{\rm{in}}$, $3R_{\rm{in}}$-$7R_{\rm{in}}$, $7R_{\rm{in}}$-$12R_{\rm{in}}$, $12R_{\rm{in}}$-$20R_{\rm{in}}$, $20R_{\rm{in}}$-$30R_{\rm{in}}$, $30R_{\rm{in}}$-$50R_{\rm{in}}$, $50R_{\rm{in}}$-$100R_{\rm{in}}$ and $100R_{\rm{in}}$-$200R_{\rm{in}}$). Panels (a) to (f) correspond to situations with jet velocities $0c$, $0.1c$, $0.2c$, $0.4c$, $0.7c$, $0.99c$, respectively.}
    \label{difposition_D}
\end{figure}

To calculate the reflection spectrum, it is first necessary to carefully analyze the photons scattered back to the disk. These photons serve as the source of the reflection component, and their spectra change along the radius of the disk. We divide the accretion disk into eight regions: $1R_{\rm{in}}$-$3R_{\rm{in}}$, $3R_{\rm{in}}$-$7R_{\rm{in}}$, $7R_{\rm{in}}$-$12R_{\rm{in}}$, $12R_{\rm{in}}$-$20R_{\rm{in}}$, $20R_{\rm{in}}$-$30R_{\rm{in}}$, $30R_{\rm{in}}$-$50R_{\rm{in}}$, $50R_{\rm{in}}$-$100R_{\rm{in}}$ and $100R_{\rm{in}}$-$200R_{\rm{in}}$. For each of these five regions, we collect the returning photons to accumulate an energy spectrum, as shown in Figure \ref{difposition_D}.

The reflected spectrum is obtained by convolving each accumulated energy spectrum with the \textbf{xilconv} model in XSPEC. This model, initially proposed by \cite{Done2006}, is an updated version of the \textbf{rfxconv} model \citep{Kolehmainen2011}. The key difference lies in the reflection tables: \textbf{rfxconv} employs tables generated by the \textbf{reflionx} code \citep{Ross1999,Ross2005}, whereas \textbf{xilconv} adopts those produced by \textbf{xillver}. The computational workflow of \textbf{xilconv} involves determining the average power-law index of the input continuum spectrum within the 2–10 keV energy band and selecting the corresponding reflection spectrum from precomputed tables \citep{Garc2022}. This reflection model is chosen for several reasons. First, \textbf{xilconv} carefully calculates the physical processes the incident photons may experience in the disk and figures out the reflection spectrum at a given inclination angle, but it does not care about the position or angular distribution of the incident photons. In other words, this model can be adapted to any corona geometry, and our radius-divided data processing described in the last paragraph can supplement the geometric information into the reflection calculation. Second, unlike some other reflection models that require some spectral parameters as input, \textbf{xilconv} is a convolution model and it can accept any numerical input spectrum, so that the MC results can be used directly. Third, this model does not take the general relativistic effects into consideration, which is the case in our MC simulation program. Introducing GR effects selectively (e.g., in reflection calculations but not in photon scattering) would lead to theoretical inconsistencies. In general, the \textbf{xilconv} model is rough but acceptable, because this work is focused on a weak reflection situation (see Section \ref{sec:disc}
 for detailed discussions).

\begin{figure}[htbp!]
    \centering
    \includegraphics[width=0.75\linewidth]{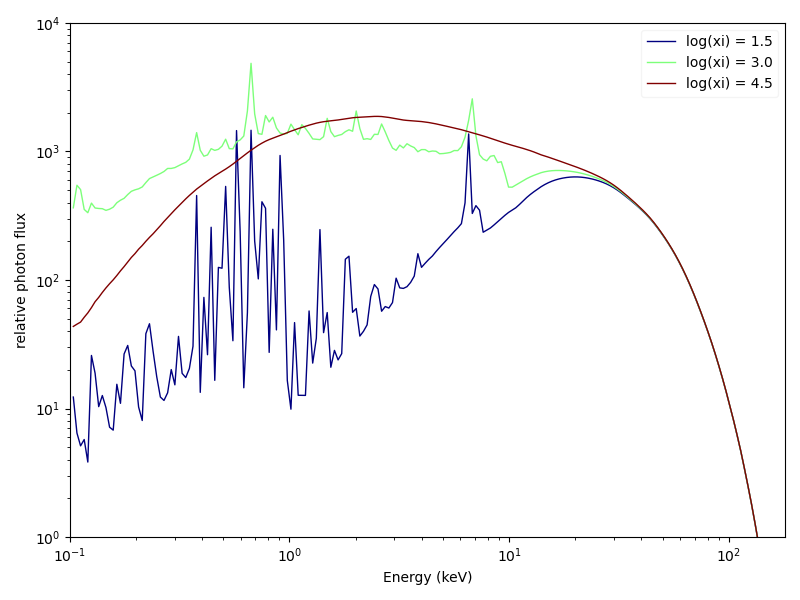}
    \caption{The reflection component spectra under different ionization parameters log(xi). The ionization parameter has little effect on the high energy part of the reflection component spectra.}
    \label{dif_ion}
\end{figure}

When using \textbf{xilconv}, we set the abundance of Fe elements in the accretion disk to the solar abundance. In addition, considering that both the shape and strength of the reflection spectrum depend on the ionization parameter \citep{zycki1994}, we test the reflection component spectra under different ionization parameters, as shown in Figure \ref{dif_ion}. The input spectrum  which is randomly picked from the spectra of return-to-disk photons in the different radial bins here is the energy spectrum of photons scattered back to the disk under a coronal velocity of $0.4c$, covering the region from $12R_{\rm{in}}$ to $20R_{\rm{in}}$. It can be found that the ionization parameter primarily affects the shape and intensity of the reflection spectrum in the low energy range, and the reflection spectrum at $> 20$ keV remains unchanged under different ionization degrees. The low energy part of overall spectrum is dominated by the thermal component, therefore, variations in the reflection component do not significantly affect the overall spectrum. In the subsequent sections, to avoid the underestimation of the reflection flux, we set the ionization parameter log(Xi) uniformly to 4.5 to simulate the partially ionized accretion disk(we also present the simulation results of log(Xi)=3 in Appendix \ref{appendix B}, in which emission lines are obvious but the continuum spectra are weaker). To get the parameter of the exponential cut-off energy required by \textbf{xilconv}, we use the \textbf{cutoffpl} model to fit the spectra of return-to-disk photons shown in Figure \ref{difposition_D} (see details in Appendix \ref{xilconv cut-off}). Finally, reflections from different radial regions of the disk are combined to obtain the observed reflection spectra at inclinations of 25°, 35°, 45°, ... 85°.
% \begin{figure}[htbp!]
%     \centering
%     \includegraphics[width=1.0\linewidth]{input_cutoffpl.png}
%     \caption{The left panel shows comparisons between a cutoffpl (red) and our input spectrum (blue). The right panel shows the reflection spectra obtained by convolving both the cutoff power law spectrum and the return-to-disk photons spectrum with the xilconv model.}
%     \label{input_cutoffpl}
% \end{figure}

\section{simulation results} \label{sec:results}

\subsection{the return-to-disk photons} \label{returning disk component}

\begin{figure}[htbp!]
    \centering
    \includegraphics[width=0.75\linewidth]{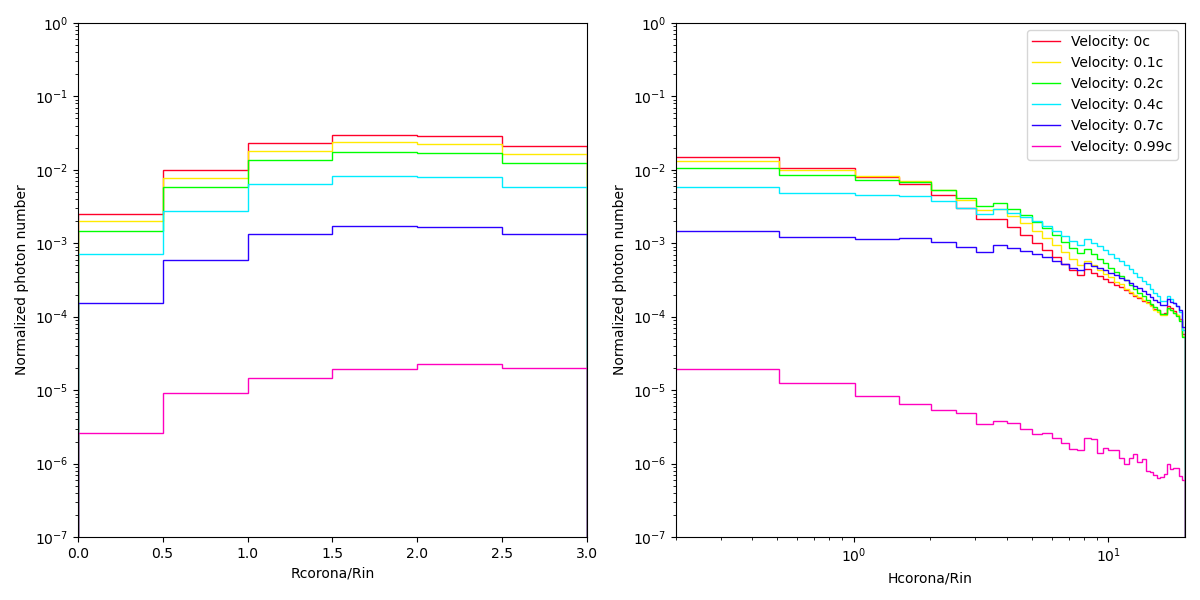}
    \caption{The densities of positions where return-to-disk photons exit the cylindrical corona on the surface of the corona. The left panel shows the densities at different radii on the bottom side of the cylindrical corona. The right panel shows the densities at different height on the lateral side of the cylindrical corona. The photon number is normalized to the seed photon number. The lines with different colors correspond to simulations with different jet velocities.}
    \label{fig2}
\end{figure}
Current reflection models calculate the reflection spectrum by customizing the irradiation spectrum to the accretion disk. However, the emission position of scattered photons that illuminate the disk and the location where they reach the disk also influence the spectral features of the reflection spectrum. Therefore, in addition to counting the energy spectrum of the return-to-disk photons, we also make statistics on the positions where these photons leave the corona and where they reach the accretion disk. Here, we plot the distribution of positions on the surface of the jet corona where the scattered photons exit the corona, as shown in Figure \ref{fig2}. It can be seen that for a cylindrical corona with radius of $3R_{\rm{in}}$, most of the return-to-disk photons emerge from the bottom side of the corona, concentrating around a radius of $2R_{\rm{in}}$. The number of photons emerging from the lateral side of the corona decreases with the increasing height.
The fact that most return-to-disk photons are emitted from the bottom side of the corona indicates that when considering the reflection radiation between the jet-like corona and the disk, the radius of the corona, i.e., the size of the inner part of the disk that is covered by the corona, is more significant than the shape of the corona. Therefore, modeling the jet base with a cylindrical shape is a reasonable way to simplify the geometry.

\begin{figure}[htbp!]
    \centering
    \includegraphics[width=0.75\linewidth]{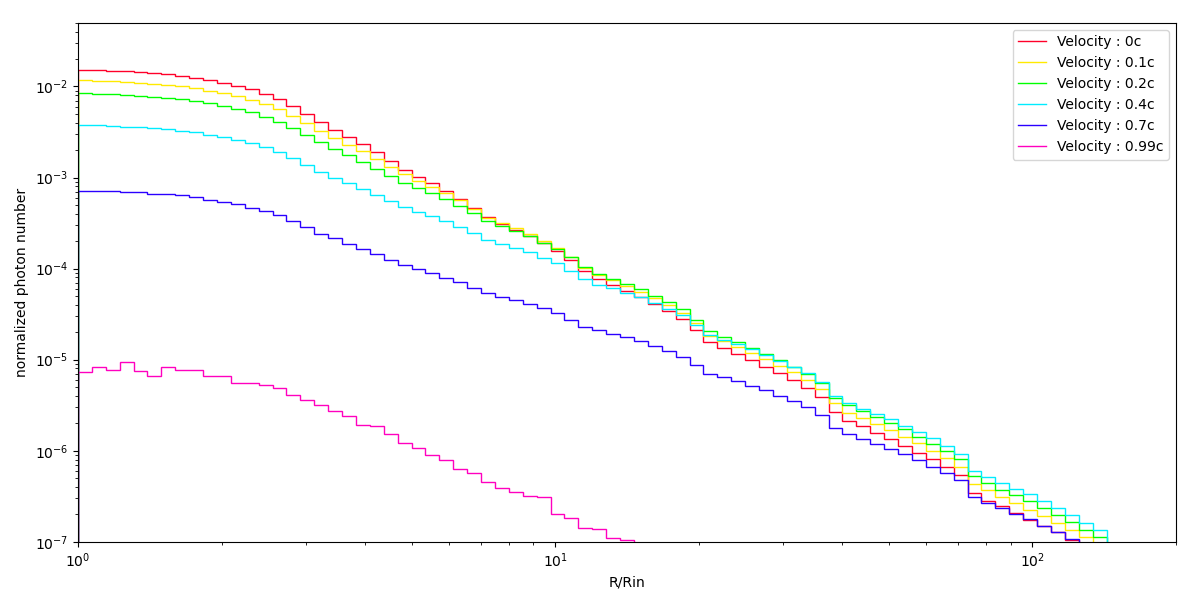}
    \caption{The densities of positions where return-to-disk photon reaches the accretion disk. The photon number is normalized to the seed photon number. The lines with different colors correspond to simulations with different jet velocities.}
    \label{D_position_density}
\end{figure}

% We also plot the two-dimensional probability density distribution of the positions where the return-to-disk photons reach the accretion disk, as shown in Figure \ref{fig3}.
We also plot the density distribution of the positions where the return-to-disk photons reach the accretion disk, as shown in Figure \ref{D_position_density}.
It is shown that the probability density decreases quickly when the radius increases, indicating that the reflection component is primarily generated in the region nearby the inner radius of the disk. It should be noticed that this result is based on a geometry that the vertical distance of the corona to the disk ($1R_{\rm{in}}$) is smaller than its radius ($3R_{\rm{in}}$); for a more distant corona, the illuminated area on the disk will be more extended. When the jet velocity is extremely large, the disk receives very little radiation from the corona, due to the special relativistic effect.

% \begin{figure}[htbp!]
%     \centering
%     \includegraphics[width=0.75\linewidth]{position of disk.png}
%     \caption{Probability densities of the positions where a return-to-disk photon reaches the accretion disk. Panels (a) to (f) correspond to situations with jet velocities $0 c$, $0.1 c$, $0.2 c$, $0.4 c$, $0.7 c$, $0.99 c$, respectively.}
%     \label{fig3}
% \end{figure}

\subsection{the Spectral results} \label{subsec:overall energy spectra}
\begin{figure}[htbp!]
    \centering
    \includegraphics[width=0.75\linewidth]{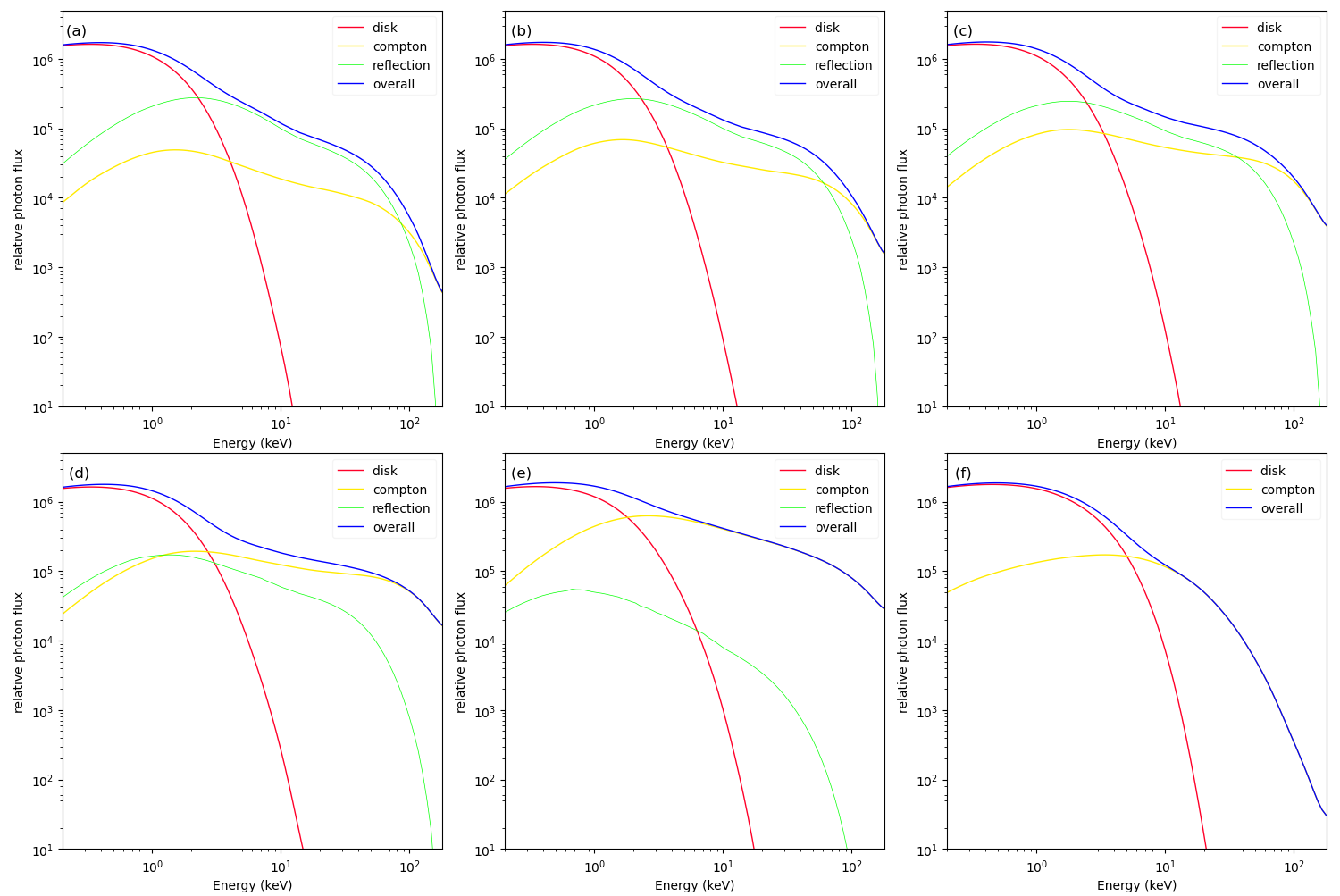}
    \caption{The simulated spectra at inclination angle 25°. The red lines are the thermal component emitted by the accretion disk; the yellow lines are the Compton component; the green lines are the reflection component; the blue lines are the sum of all components, i.e., the observed spectra. Panels (a) to (f) correspond to jet velocity $0 c$, $0.1 c$, $0.2 c$, $0.4 c$, $0.7 c$, $0.99 c$, respectively.}
    \label{fig4}
\end{figure}

\begin{figure}[htbp!]
    \centering
    \includegraphics[width=0.75\linewidth]{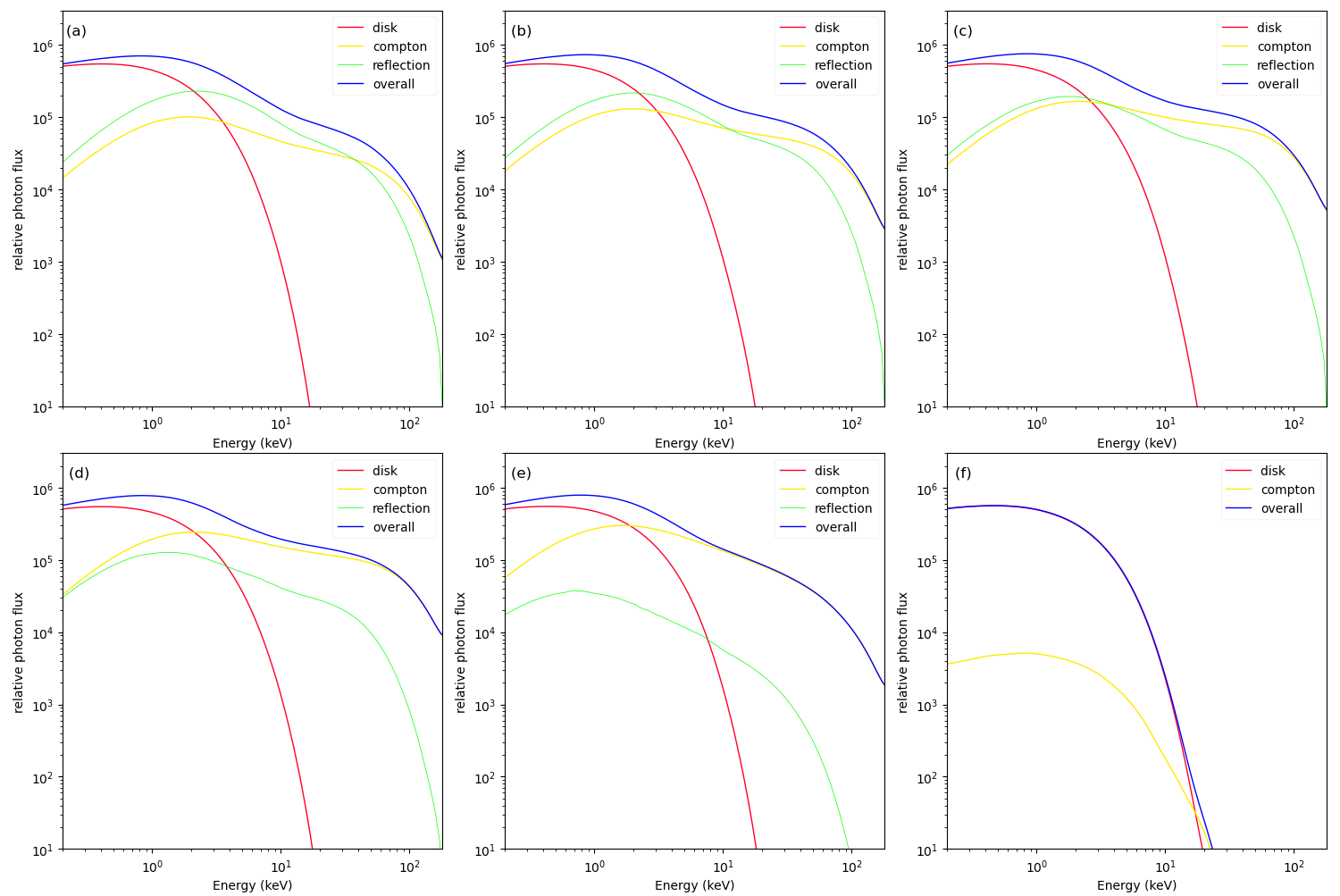}
    \caption{The simulated spectra at inclination angle 55°. The red lines are the thermal component emitted by the accretion disk; the yellow lines are the Compton component; the green lines are the reflection component; the blue lines are the sum of all components, i.e., the observed spectra. Panels (a) to (f) correspond to jet velocity $0 c$, $0.1 c$, $0.2 c$, $0.4 c$, $0.7 c$, $0.99 c$, respectively.}
    \label{fig5}
\end{figure}

\begin{figure}[htbp!]
    \centering
    \includegraphics[width=0.75\linewidth]{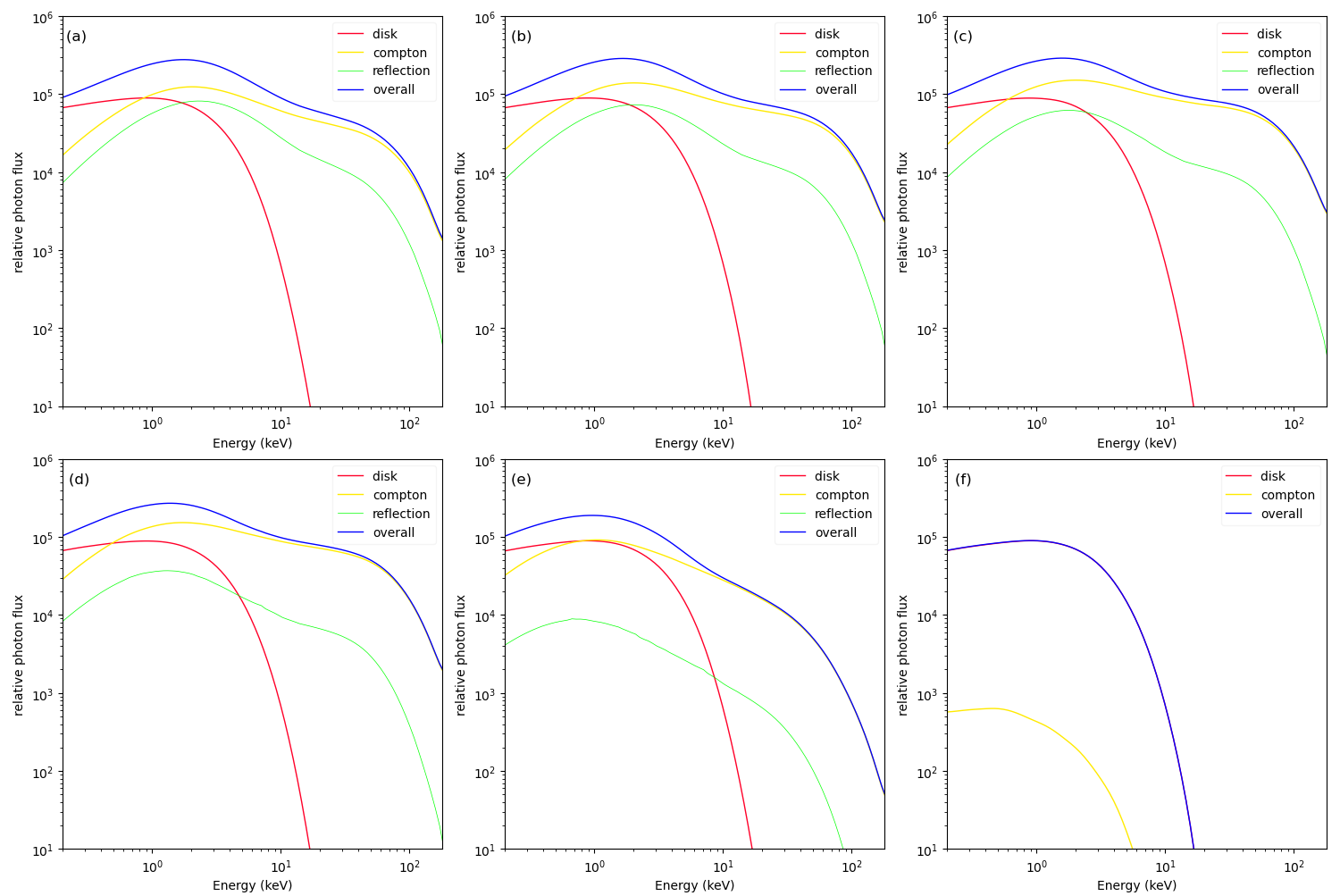}
    \caption{The simulated spectra at inclination angle 85°. The red lines are the thermal component emitted by the accretion disk; the yellow lines are the Compton component; the green lines are the reflection component; the blue lines are the sum of all components, i.e., the observed spectra. Panels (a) to (f) correspond to jet velocity $0 c$, $0.1 c$, $0.2 c$, $0.4 c$, $0.7 c$, $0.99 c$, respectively.}
    \label{fig6}
\end{figure}

Figures \ref{fig4}-\ref{fig6} are the overall simulated spectra and spectral components for inclination angles in sequence of 25°, 55°, and 85° (the results for other inclination angles are shown in Appendix \ref{appendix A}); for each inclination angle, the spectra of different jet velocities are plotted separately in different panels. In addition, we also include the energy spectra corresponding to log(xi)=3.0 in Appendix \ref{appendix B}. Except for the extreme velocity of $0.99 c$, the spectra exhibit a double-hump feature. The low energy hump originates from the multicolor blackbody spectrum emitted by the accretion disk. The high energy hump is the combination of the reflection component and the Compton component. In general, it is the Compton component that dominates the high energy hump, especially for energies $> 100$ keV where almost all radiation is contributed by the Compton component. The flux of the reflection component could be higher than the Compton component in the middle energy range only when both the inclination angle and the velocity of the jet are small. The spectrum of the scattering component is significantly affected by the temperature and the optical depth of the corona  \citep{Kompaneets1957,Poutanen1996,Zdziarski2020} and Compton up-scattering can cause an enhancement of flux in the high energy band. The coronal temperature $T_{\rm{e}}$ determines the energy range of the Compton hump, with higher temperatures enabling photons to gain more energy per scattering, thereby shifting the peak of the non-thermal hump to higher energies. The scattering optical depth $\tau$ controls the shape of the Compton hump, as a larger optical depth increases the average scattering number of photons, leading to a wider hump or even a Compton plateau.

\begin{figure}[htbp!]
    \centering
    \includegraphics[width=0.75\linewidth]{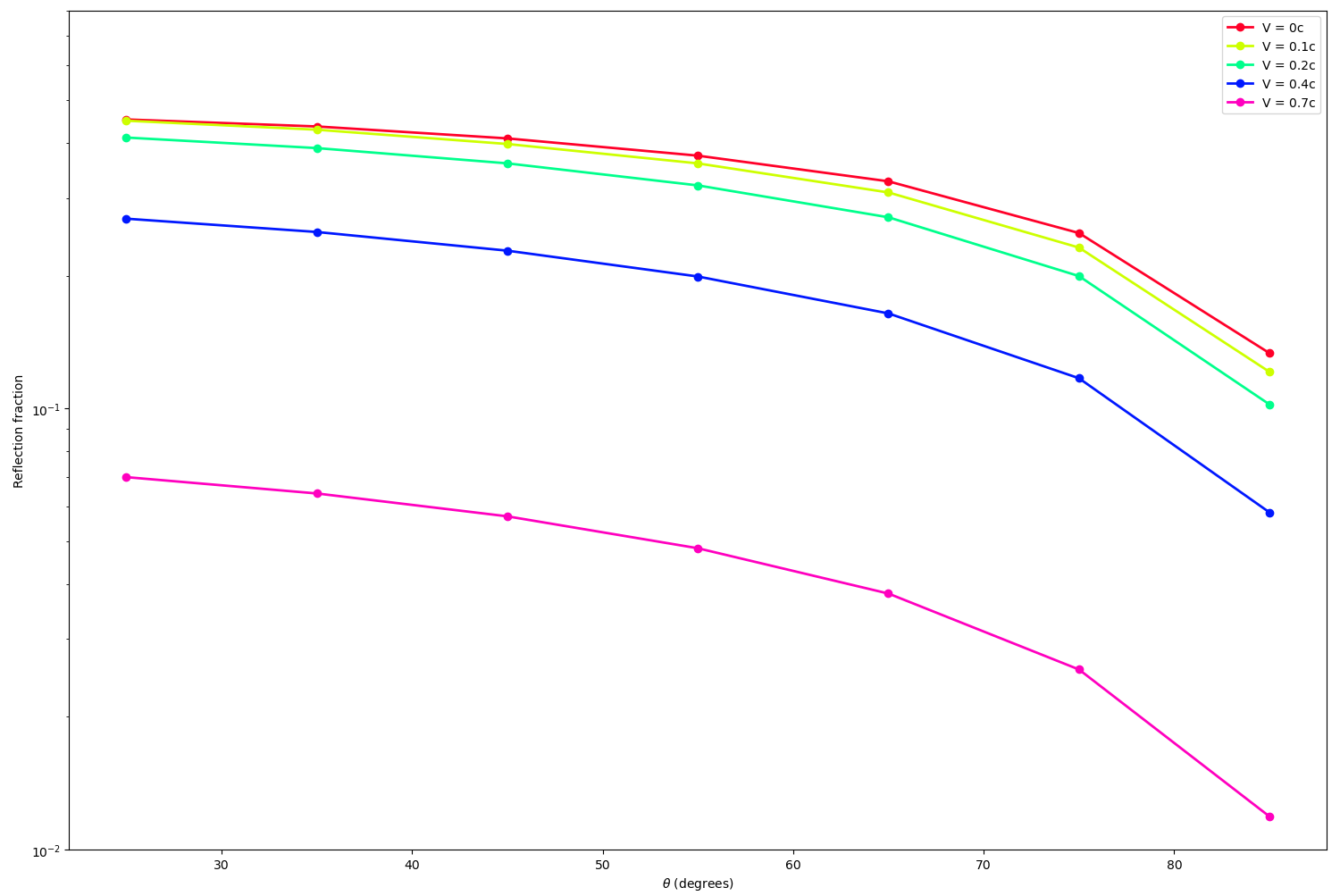}
    \caption{Variation of reflection fraction with inclinations. The lines with different colors correspond to simulations with different jet velocities.}
    \label{ref_fraction}
\end{figure}

In addition to the double-hump feature of the overall spectra, the simulation results also reveal the notable impact of the jet velocity on the spectra.
Figure \ref{ref_fraction} shows the reflection fraction across different coronal velocities and inclination angles.
We find that as the jet velocity increases, the proportion of the reflection component decreases rapidly. For all observing inclination angles, the reflection component is negligible if the jet velocity is above $0.4 c$; in this case, almost all the high energy radiation is contributed by the Compton component. If the jet is accelerated to near the velocity of light, the final orientations of photons in the Compton component will be limited in small inclination angles, so that observers at large inclination angles can only see the disk component.

\section{Discussion} \label{sec:disc}

In Section \ref{sec:results}, we made multiple simulations under a scenario that the jet near the BH serves as a Compton scattering zone, simplified as a cylindrical corona; and all the seed photons are generated by the thermal radiation of the accretion disk. We refer to this scenario as the disk-seed jet-scattering scenario.

Our simulations are conducted in flat spacetime; therefore, the simulated spectra are not quite accurate due to the lack of General Relativistic (GR) effects. GR effects can exert significant influences on the disk reflection: light bending affects the radial distribution of the incident photons of the disk reflection, and the relativistic blurring changes the shape of the disk emission lines. However, for the Comptonization process in the jet, the GR effects as a whole do not result in qualitative differences because the gravitational field is weak at large distance to the central BH, as shown in Figure \ref{geant4_monk}. On the other hand, as described in Section \ref{sec:results}, the flux of the reflection component is much smaller than the Compton component at hard energy range. Consequently, despite of the lack of GR effects calculations, our simulations are reliable in terms of weak reflection situations and Comptonization process. This means that GR effects do not qualitatively affect our conclusion that a weak reflection component is produced for a jet-like corona. In the following paragraphs, we will make discussions on the mechanisms of disk reflection weakened by the relativistic jet and the source of the Compton component.

In our simulations, the model geometry is chosen to be a homogeneous cylinder for simplification. The primary distinction between a cylinder and a sphere lies in the direction perpendicular to the disk: the cylindrical corona exhibits a significantly larger vertical optical depth. Thus, even at high coronal velocities, photons have a high probability of undergoing scattering rather than traversing the corona. This makes the cylindrical corona more suitable to the phenomenon in which the reflection is weak meanwhile the Compton component is strong. To clearly show the difference between spherical geometry and cylindrical geometry, a spherical corona with the same temperature, height and radial optical depth is simulated and compared with the cylindrical corona in panel (a) of Figure \ref{geant4_monk}. It is obvious that the Compton component of the cylindrical corona is stronger than that of spherical corona, which is expected.
In addition to the corona geometry, the radius and height of the corona (which represents the size of the jet) are fixed at a moderate group of values. It is obvious that these parameters can have a significant influence on the scattered ratio of the seed disk photons: a larger corona radius $R_{\rm{corona}}$ results in a higher scattering rate of seed photons, whereas a greater corona height $h_{\rm{corona}}$ leads to a lower scattering rate of seed photons. Crudely, the solid angle of the jet in the view of the inner part of the disk decides the scattered ratio of the seed disk photons. This is because most of the seed photons are emitted from the inner part of the disk, and the larger the solid angle is, the more the seed photons are radiated into the jet base. On the other hand, the velocity of the jet has little influence on the scattered ratio of photons in the jet. The jet is above the disk and the size of the jet base is larger than the inner radius of the disk so that the column density of the jet on the view direction of the inner part of the disk is large enough to counteract the reduction in scattering cross-section caused by the relativistic effects of the jet motion. Therefore, the intensity of the Compton scattering component is only slightly affected by the velocity of the jet.

\begin{figure}[htbp!]
    \centering
    \includegraphics[width=0.75\linewidth]{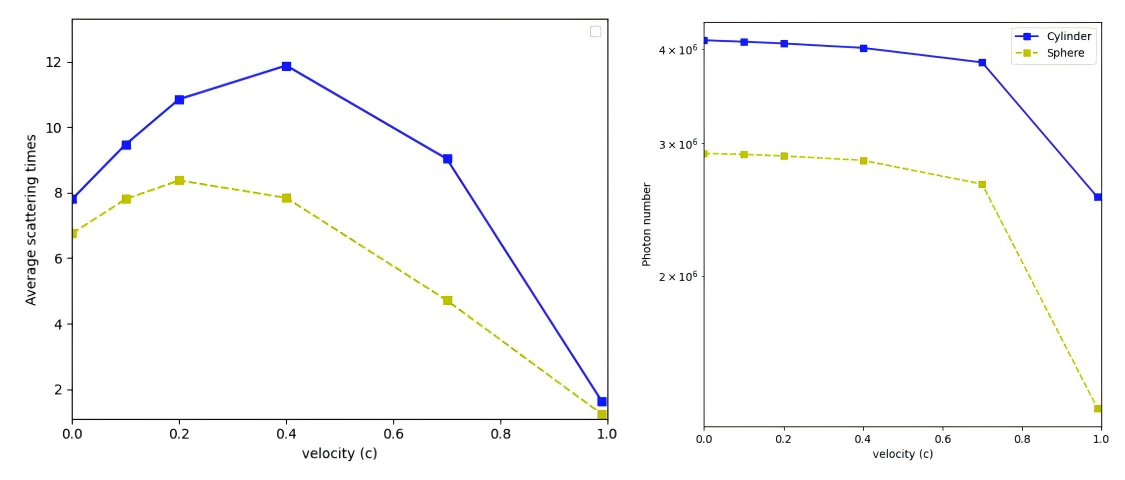}
    \caption{The average scattering times and the total scattered photon counts between the cylindrical corona (blue solid line) and the spherical corona (yellow dashed line) under different jet velocities.}
    \label{Sphere-Tubs}
\end{figure}

On the contrary, the intensity of the disk reflection component in the observed spectrum is strongly dependent on the velocity of the jet. According to the simulation results shown in Section \ref{returning disk component}, scattered photons which illuminate the disk are mainly emitted from the base region of the jet. If the position of the jet base and the size of the jet and disk are fixed, then the solid angle of the disk as viewed by the jet base is constant regardless of whether the jet is slow or fast. In this case, the decisive factor is the directional distribution of scattered photons, which is deeply affected by the velocity of the jet: in a relativistic jet, photons are more likely to be scattered toward the direction of the bulk motion of the electrons, which is away from the disk; therefore, the disk will not be illuminated effectively by the scattered photons if the bulk motion is relativistic.
\begin{figure}[ht!]
    \centering
    \includegraphics[width=0.75\linewidth]{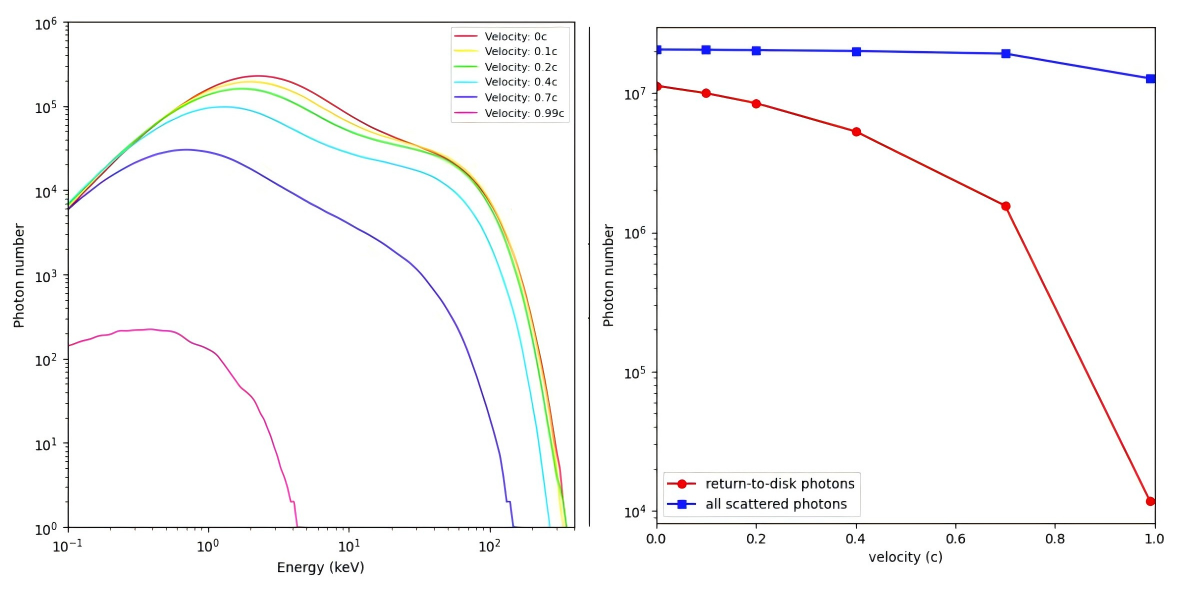}
    \caption{The influences of the jet velocity on return-to-disk photons. The left panel shows the spectra of return-to-disk photons under different jet velocities. The right panel shows the number of return-to-disk photons (red line) under different jet velocities, compared with the number of all scattered photons (blue line). }
    \label{fig7}
\end{figure}

Figure \ref{fig7} shows the energy spectra and the total number of the return-to-disk photons when the coronal velocity is $0 c$, $0.1 c$, $0.2 c$, $0.4 c$, $0.7 c$, $0.99 c$, respectively. The peak of the energy spectrum shifts to lower energies as the jet velocity increases. The number of the return-to-disk photons decreases as the jet velocity increases because the directions of scattered photons in a jet with higher velocity tend to be more aligned with the jet direction, i.e., the direction away from the disk. Meanwhile, the number of all scattered photons changes little with the jet velocity, except for the slight decrease when the velocity of the jet is extreme relativistic. Hence, the jet velocity has few influences on the scattered ratio of seed photons. As a result, the velocity of the jet has a much larger influence on the flux of the reflection component than on the Compton component.

\begin{figure}[ht!]
    \centering
    \includegraphics[width=0.75\linewidth]{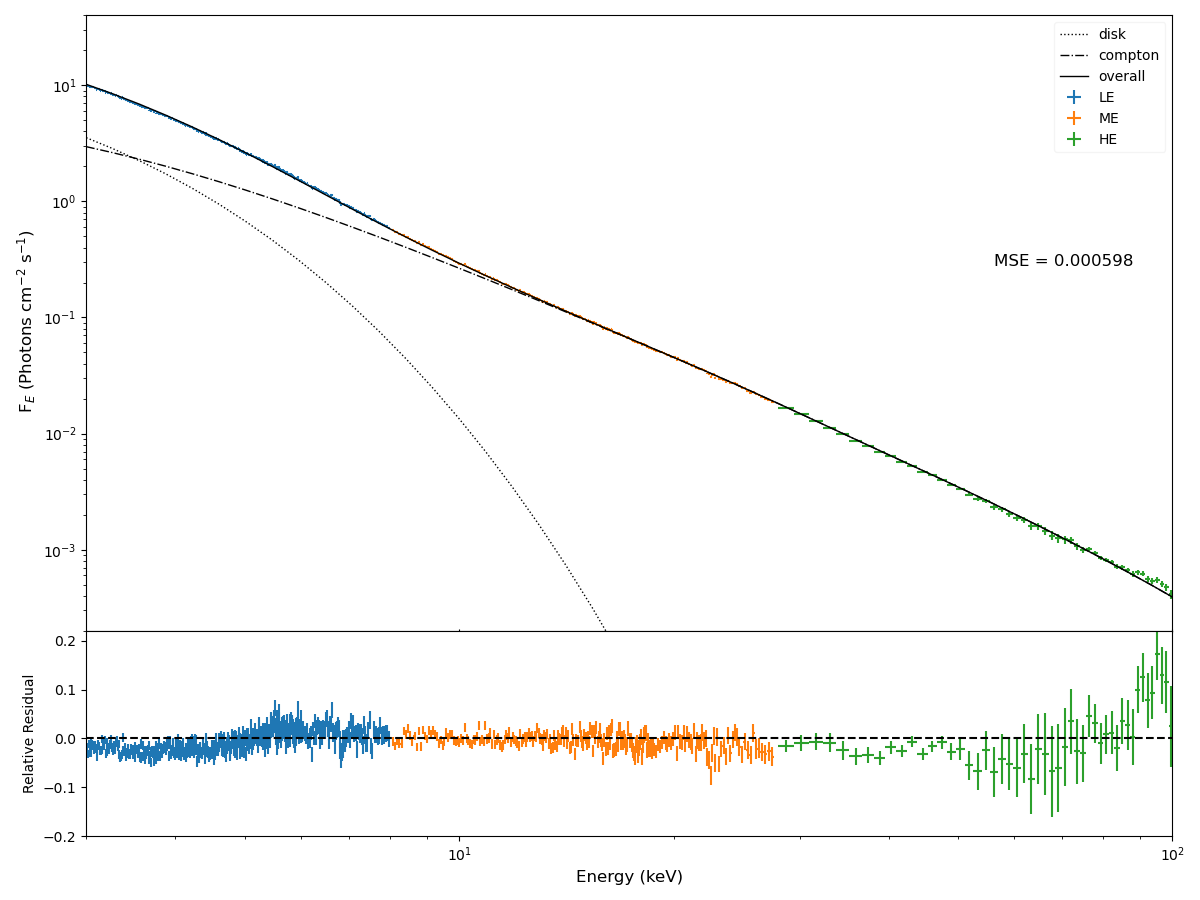}
    \caption{The unfolded spectrum of Swift J1727.8-1613 observed by Insight-HXMT compared with MC simulation results; the lower panel is the corresponding relative residual. The selected observation is NO. P061433802003, which was during VHS of its 2023 burst. The blue, orange, and green curves represent the low-energy (LE), medium-energy (ME), and high-energy (HE) data, respectively; the solid line, dash-dotted line and the dotted line denote the MC spectrum, its Compton component and its disk component, respectively. The mean squared error (MSE) of the log-transformed data is 0.000598 and the corresponding MC parameters are shown in Table \ref{parameters}.}
    \label{fit_VHS}
\end{figure}

These very different impacts of the jet velocity on the Compton component and the reflection component lead to an interesting inference: in a jet-corona scenario with all seed photons generated by the disk, a weak disk reflection component and a strong Compton component at high energy range can coexist in the observed spectrum. In other words, our simulation results demonstrate that the possibility of the accretion disk offering seed photons for scattering cannot be ruled out by the absence of the reflection. This is actually the case of Swift J1727.8-1613, a BH X-ray binary discovered on August 24, 2023 \citep{Castro-Tirado}. A possible jet/corona configuration is adjusted to account for the spectral fitting with different model trials, and the absence of the reflection component in the spectrum indicates that the jet has a large velocity \citep{Peng, Cao_swift1727}. To test the disk-seed jet-scattering scenario on Swift J1727.8-1613, observation data and MC simulation results are compared. The Insight-HXMT observation NO. P061433802003 (MJD 60206.27) is selected, which was during the very high state (VHS) flaring period of its 2023 burst. This HXMT data is not the intrinsic radiation of the observed X-ray source, but the convolution result of the source radiation and the response of HXMT detectors. Therefore, to exclude the impact of the response of the detectors, a spectral model is needed: the observed data is fitted with the spectral model folded with the detector response, then the properties of the source are reflected by the best fit model parameters. Since our Geant4 simulation results have not been summarized as a table model to be used in spectral fitting, we have to select a spectral model to take the place of our Geant4 simulation when making the spectral fitting. We choose a XSPEC model \textbf{constant $\times$ tbabs $\times$ (diskbb + relxill + powerlaw)}\citep{Cao_swift1727} which conforms to our Geant4 simulation settings in general\footnote{The difference between Cao's model and our model is that there are two coronas in Cao's model, i.e. the corona illuminating the disk and the corona generating the scattering component in the spectrum are two different coronas.}, and fit the data with the XSPEC model to get an unfolded spectrum (net spectrum) of the source. Then a visual match is done between the unfolded spectrum and our Geant4 simulation result by manually tuning the MC parameters. The XSPEC model parameters used in getting the unfolded spectrum are shown in Table \ref{model component parameters}. The visual match is shown in Figure \ref{fit_VHS}, with the corresponding MC parameters shown in Table \ref{parameters}. The reflection component is weak enough to be ignored in our MC simulation, which can be explained by the large velocity and the small scale of the jet, and this weak reflection inference is consistent with the XSPEC fitting result. The unfolded spectrum and the MC spectrum are in line with each other in general, indicating that this source likely had a relativistic jet during the VHS flaring period, Comptonizing the disk radiation and offering weak illumination back to the disk. However, due to the absence of GR effects at present, the parameter values of this MC simulation may not be accurate.

\begin{table}[htbp!]
\centering
\caption{XSPEC model parameters of the unfolded spectrum of Insight-HXMT observation NO. P061433802003 of Swift J1727.8-1613.}
\begin{tabular}{ccc}
\toprule %表格顶部粗横线
Component & Parameters & Value \\
\hline
\multirow{4}{*}{constant}& $factor1$ &    \multirow{2}{*} {0.911$\pm  4.385× 10^{-3}$}\\
& (scaling factor for Insight-HXMT/ME data) \\
& $factor2$ &    \multirow{2}{*} {0.888$\pm  9.604× 10^{-3}$}\\
& (scaling factor for Insight-HXMT/HE data)\\
\hline
\multirow{2}{*}{tbabs}& $N_{\rm{H}}$  &  \multirow{2}{*}{  $0.4 × 10^{22}\  \rm{atom\ cm^{-2}}$$^{\star}$}\\
&(equivalent hydrogen column) \\
\hline
\multirow{3}{*}{diskbb}& $T_{\rm{in}} $ &  \multirow{2}{*}  {1.121$\pm$$2.227× 10^{-3}\ \rm{keV}$}\\
&(temperature at inner disk radius) \\
& $norm$ &    4614.59$\pm$60.61\\
\hline
\multirow{22}{*}{relxill}& $Index1{\&}Index2$  &  \multirow{2}{*}{  3$ ^{\star}$}\\
& (the emissivity indices) \\
& $R_{\rm{in}}$ &    \multirow{2}{*} {1 $\rm{ISCO}$ $ ^{\star}$}\\
& (inner radius of the accretion disk ) \\
& $R_{\rm{out}}$  &    \multirow{2}{*} {400$R_{\rm{g}}$$ ^{\star}$}\\
& (outer radius of the accretion disk) \\
& $a$ &  \multirow{2}{*} {0.998$ ^{\star}$}  \\
& (spin of black hole) \\
& $Incl$ &  \multirow{2}{*} {47.9°$ ^{\star}$}  \\
& (inclination toward the system) \\
& $z$ &    \multirow{2}{*} {0$ ^{\star}$}\\
& (redshift to the source) \\
& $log \xi$ &   \multirow{2}{*} {3.4$ ^{\star}$} \\
& (ionization of the accretion disk) \\
& $A_{\rm{fe}}$ &  \multirow{2}{*} {0.5$ ^{\star}$}  \\
& (the iron abundance of the material in the accretion disk) \\
& $\rm{Gamma}_{\rm{rel}}$ &    \multirow{2}{*} {1.883$\pm$0.0486}\\
& (power-law photon index of the incident spectrum, $E^{\rm{-Gamma}}$) \\
& ${E}_{\rm{cut}}$ &   \multirow{2}{*} {20.7126$\pm$1.3581$\ \rm{keV}$} \\
& (high-energy cutoff of the primary spectrum) \\
& $f_{\rm{ref}}$ &    \multirow{2}{*} {$0.0032612^{+2.71}_{-0.0032612}× 10^{-2}$}\\
& (reflection fraction) \\
& $norm$ &    0.1319$\pm$0.0152\\
\hline
\multirow{3}{*}{powerlaw}& $\rm{Gamma}_{\rm{pl}}$ &  \multirow{2}{*} {2.6$\pm$0.017}\\
& (photon index of power law, $E^{\rm{-Gamma}}$)\\
& $norm$ &    65.1274$\pm$1.4893\\

\hline
Note. $ ^{\star}$ means the parameter is fixed.
\label{model component parameters}
\end{tabular}
\end{table}

\begin{table}[htbp!]
\centering
\caption{The approximately estimated parameters of the MC simualtion corresponding to Insight-HXMT observation NO. P061433802003 of Swift J1727.8-1613. No errors are available since we did not perform a spectral fitting, but only visually tuned the parameters to match the data.}
\begin{tabular}{ccc}
\toprule %表格顶部粗横线
\multicolumn{2}{c}{Parameters} & Value \\
\hline
\multirow{6}{*}{Corona}& Velocity &    0.75 $\rm c$\\
& Radius &    $1.335R_{\rm{in}}$\\
& Height  &    $20R_{\rm{in}}$\\
& Vertical distance from the black hole center &    $1R_{\rm{in}}$\\
& Optical depth &    1.3334\\
& Temperature &    $17.92\ \rm{keV}$\\
\hline
\multirow{2}{*}{Disk}& inclination &    35°\\
& Temperature &    $0.65839\ \rm{keV}$\\
\hline
\label{parameters}
\end{tabular}
\end{table}

However, there is another competitive interpretation that the Compton scattering components in their spectra are generated by the SSC scattering in the jet. Commonly the SSC process is believed to generate a (cutoff) power law spectral components with a constant photon index in a wide energy range \citep{Poutanen2014}, because the energy distribution of electrons in a jet is wide; in contrast, a Compton component with electrons of Maxwellian distribution at a certain temperature generally has a hump shape. However, the attempt to distinguish the SSC scenario and disk-seed jet-scattering scenario with the spectral shape is unreliable, because the latter can also generate smooth spectra without obvious hump structure under selected parameter combinations. In SSC, the radiation mechanisms of the disk and the jet are totally decoupled by the relativistic velocity of the jet, so that the disk component and the jet component share few common characteristics, and sometimes the disk component can be absent. On the contrary, in the disk-seed jet-scattering scenario, the Compton scattering in the jet base is coupled with the disk thermal radiation, therefore, the Compton component may inherit some characteristics from the disk component \citep{Lucchini2023}. Theoretically, the hard X-ray bands may have similar quasi-periodic oscillation frequency with the soft X-ray bands, or a hard latency of a few milliseconds might be observed. Another way to distinguish SSC and disk-seed jet-scattering scenario is to calculate the scattered efficiency of the seed photons \citep{Poutanen2014}. For the SSC scenario, the entire physical process is extremely complex; there is not a simple upper limit for the SSC radiation intensity. The synchrotron and the subsequent Compton processes extract energy from the kinetic energy of electrons in the jet, and the jet is able to afford this level of energy dissipation considering the intensity of observed hard X-ray radiation. In addition, most synchrotron photons should be scattered before leaving the jet, because the jet is optically thick. However, in the disk-seed jet-scattering scenario, the physics is quite clear. The generation rate of the seed photons is limited by the temperature and the inner radius of the accretion disk, and only a part of the seed photons can be transported to the jet-like corona. It remains to be seen if the disk blackbody emission can provide all the seed photons for the observed hard X-ray emissions.

\section{Summary} \label{sec:Summary}
The shape of the scattering corona and the source of the seed photons are two significant issues that the simple point-like lamp-post scenario cannot address. In this work, we make simulations on the X-ray radiation of an accreting BH under a disk-seed jet-scattering scenario: all the seed photons are from the thermal radiation of the accretion disk, and the scattering corona has a cylindrical shape located on the rotation axis of the disk, corresponding to the jet near the BH. This can be a physical realization of the lamp-post scenario. The seed photons obey the \textbf{diskbb} model; the Compton scattering process in the cylindrical corona (i.e., the jet) is calculated by a custom MC program based on Geant4 and the photons scattered back to the disk are counted as the input of the convolution reflection model \textbf{xilconv}. The simulation results show that a typical spectrum of this disk-seed jet-scattering scenario has a double-hump structure: the soft hump corresponds to the disk component and the hard hump corresponds to the combination of the corona Compton component and the disk reflection component.

In most cases, the reflection component is weaker than the Compton component, especially when the jet velocity or the inclination angle is large. The Compton component can still be obvious at energies $>100\ \rm{keV}$ even when the jet has a relativistic velocity, suggesting that the possibility of the disk offering seed photons cannot be ruled out by the absence of the reflection component in the observed spectra of a black hole X-ray binary. In other words, the disk-seed jet-scattering scenario is a competitive interpretation for accreting BHs with weak disk reflection and strong hard X-ray radiation, for instance, Swift J1727-1613. To diagnose whether the case is disk-seed jet-scattering or SSC for a specific black hole X-ray binary, one can try to analyze the correlation of low energy and high energy X-ray radiation (for instance, quasi-periodic oscillation correlations of these two energy bands) or make a supply efficiency analysis of the seed photons, because in the SSC scenario the low energy disk component and high energy Compton component are decoupled in radiation mechanism, while in the disk-seed jet-scattering scenario the Compton component may be modulated by the disk which generates seed photons.

So far, our Geant4 code which is publicly available on GitLab can only simulate Compton scattering in a flat spacetime and without polarization calculations. Currently we are working on incorporating the polarization and the GR effects in Kerr spacetime to our Geant4 code. All these improvements will be updated synchronously on Gitlab with our next paper. Then, we will use this improved Geant4 code to make a lamp-post Compton table model and post it in XSPEC. We hope that these works can be helpful to other researchers.

\begin{acknowledgments}
This work made use of the data from the Insight-HXMT mission, a project funded by the China National Space
Administration (CNSA) and the Chinese Academy of Sciences (CAS). We acknowledge support from the
National Natural Science Foundation of China (Grant
No. 12333007 and 12027803) and International Partnership Program of Chinese Academy of Sciences (Grant No.113111KYSB20190020).

\end{acknowledgments}

\bibliography{sample631}{}
\bibliographystyle{aasjournal}

\appendix
\counterwithin{figure}{section}

\section{Determining the outer disk radius $R_{\rm{out}}$}
\label{appendix Rout}
\begin{figure}[htbp!]
    \centering
    \includegraphics[width=0.75\linewidth]{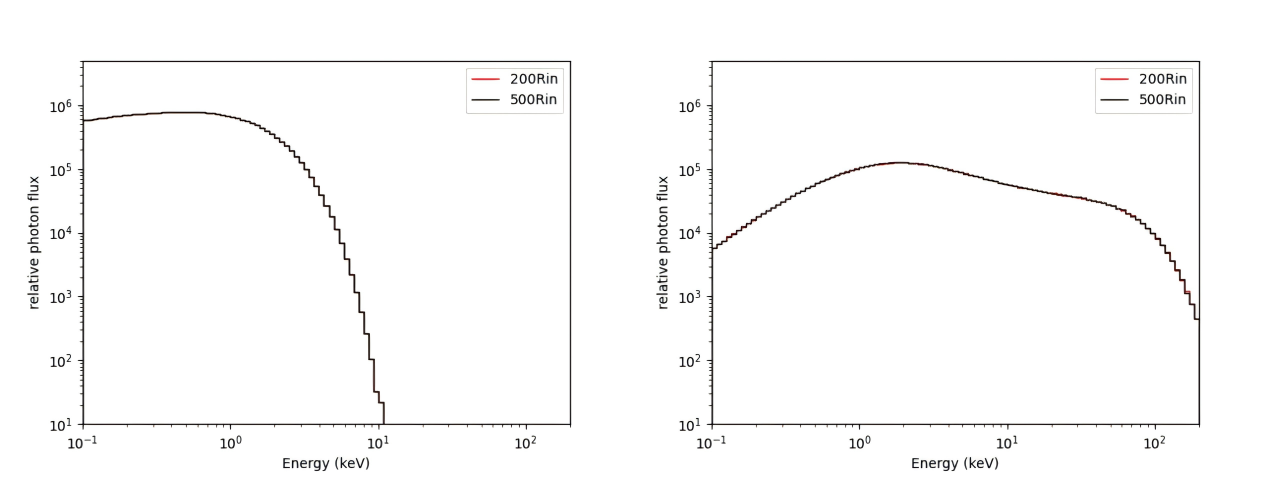}
    \caption{Comparisons of the energy spectra of disk and Compton scattering between disk outer radii of $200R_{\rm{in}}$ and $500R_{\rm{in}}$, in energy range $0.1-200$ keV. The left panel shows the energy spectra of direct disk radiation without Compton scattering with different outer disk radii. The right panel shows the energy spectra of pure Compton radiations with different outer disk radii.}
    \label{fig:difr}
\end{figure}

It is known that for the accretion disk, the larger the radius, the more photons are emitted. Therefore, simulation with larger $R_{\rm{out}}$ can generate more accurate spectral result theoretically. $R_{\rm{out}}$ is set as $200R_{\rm{in}}$ in our simulation, which needs to be tested whether this size of $R_{\rm{out}}$ is enough for the accuracy of spectrum at the concerned energy range of $0.1-200$ keV. We compared the energy spectra of the disk radiation and the Compton scattering with outer disk radii of $200R_{\rm{in}}$ and $500R_{\rm{in}}$, as presented in Figure \ref{fig:difr}. It is shown that there is no obvious difference between spectra of $200R_{\rm{in}}$ and $500R_{\rm{in}}$, both for the disk radiation and for the Compton scattering. Therefore, an outer disk radius $200R_{\rm{in}}$ is acceptable for our simulations.

\section{MC simulated spectra with log(Xi)=3}
\label{appendix B}

In this appendix, MC simulated spectra of log(Xi)=3 at inclinations 25°, 35°, 45°, ..., 75° are shown. Different from log(Xi)=4.5 where the accretion disk is highly ionized, log(Xi)=3 represent a partially ionized disk. Therefore, the reflection components of log(Xi)=3 have significant emission lines, and conversely their continuum spectra are weaker. Despite these difference, both of them share the same conclusion that the reflection component decreases quickly with the velocity of the jet, meanwhile the intensity of high-energy Compton component is slightly affected.

\begin{figure}[htbp!]
    \centering
    \includegraphics[width=0.75\linewidth]{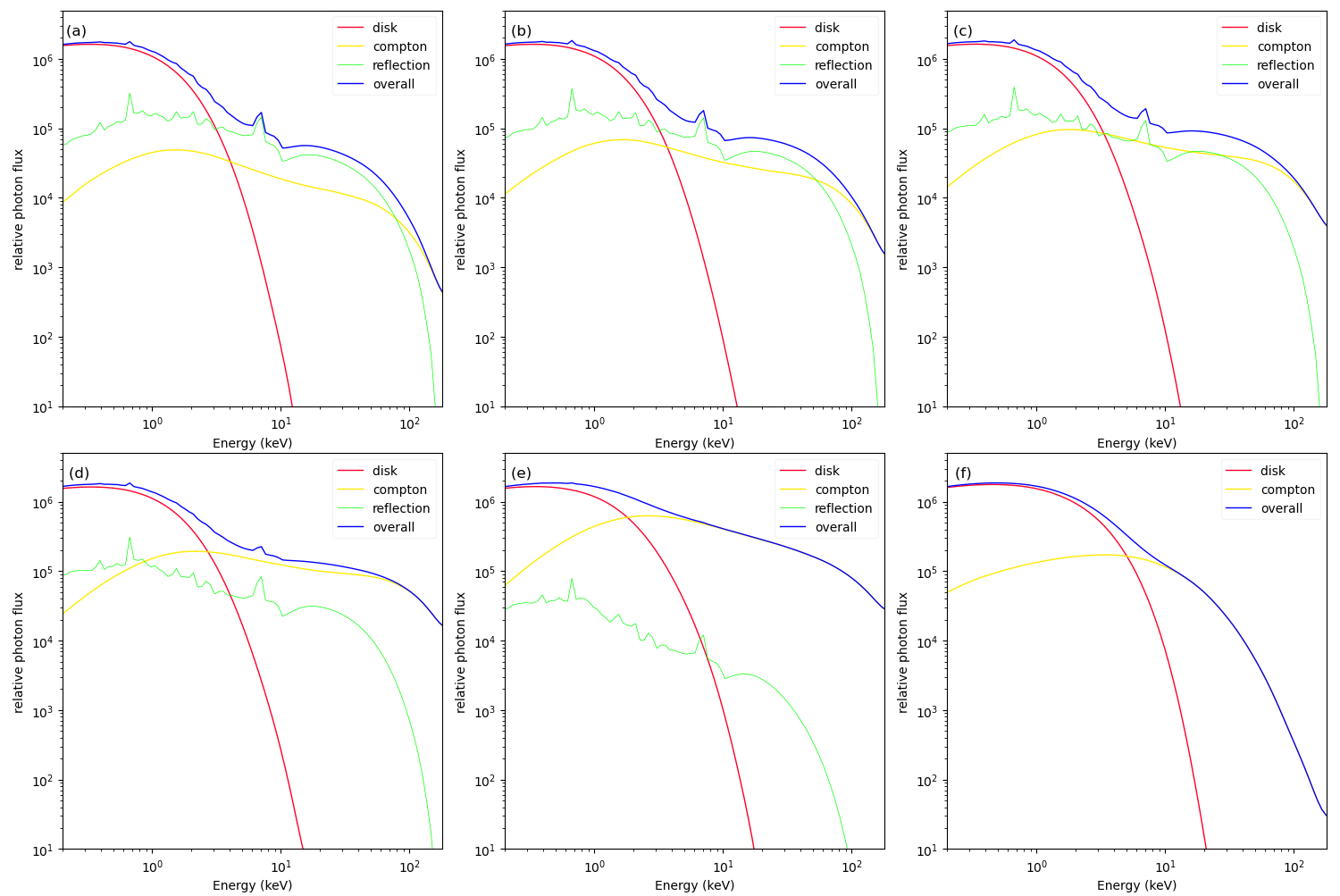}
    \caption{The simulated spectra at inclination angle 25°. The red lines are the thermal component emitted by the accretion disk; the yellow lines are the Compton component; the green lines are the reflection component; the blue lines are the sum of all components, i.e., the observed spectra. Panels (a) to (f) correspond to jet velocity $0 c$, $0.1 c$, $0.2 c$, $0.4 c$, $0.7 c$, $0.99 c$, respectively.}
\end{figure}

\begin{figure}[htbp!]
    \centering
    \includegraphics[width=0.75\linewidth]{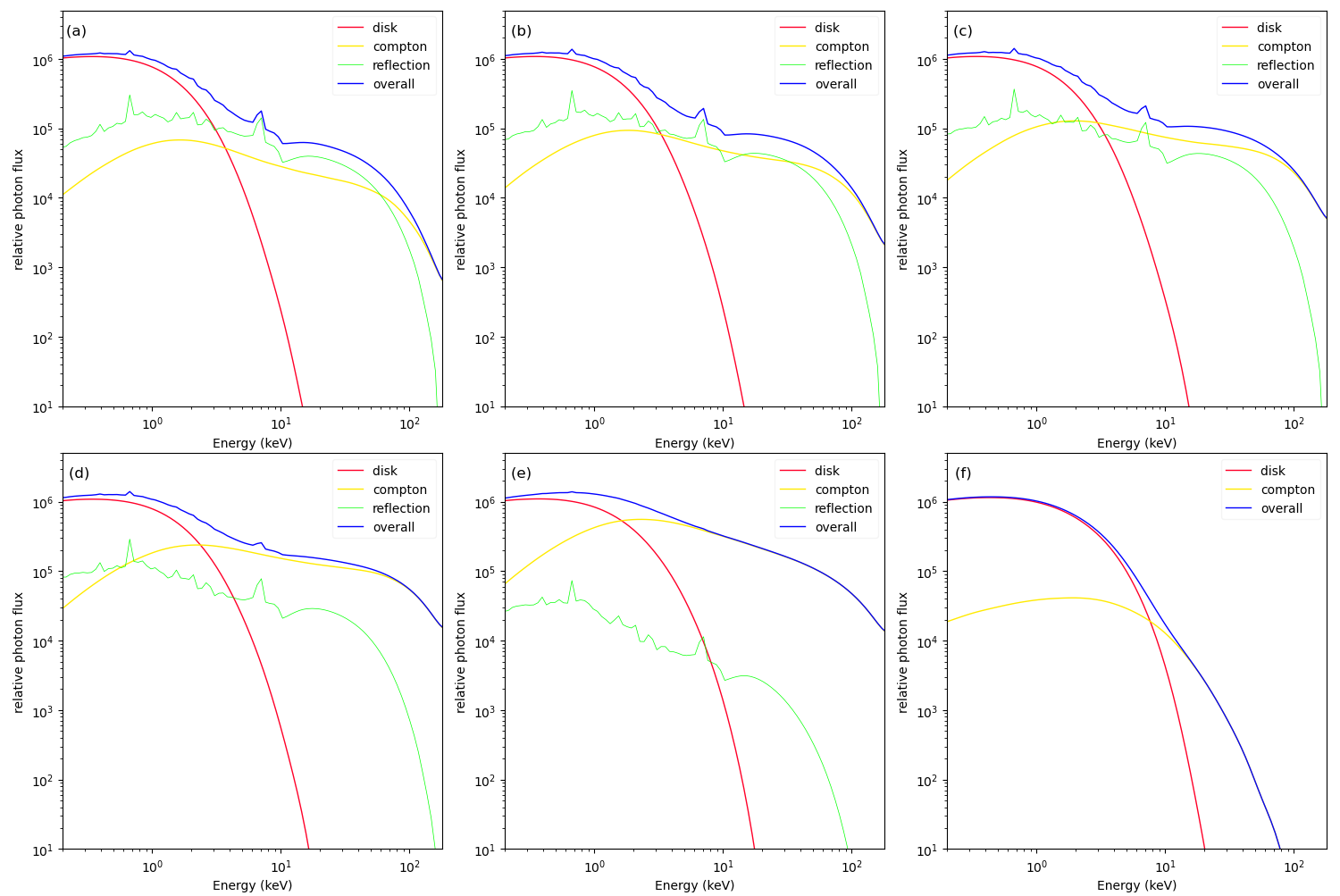}
    \caption{The simulated spectra at inclination angle 35°. The red lines are the thermal component emitted by the accretion disk; the yellow lines are the Compton component; the green lines are the reflection component; the blue lines are the sum of all components, i.e., the observed spectra. Panels (a) to (f) correspond to jet velocity $0 c$, $0.1 c$, $0.2 c$, $0.4 c$, $0.7 c$, $0.99 c$, respectively.}
\end{figure}

\begin{figure}[htbp!]
    \centering
    \includegraphics[width=0.75\linewidth]{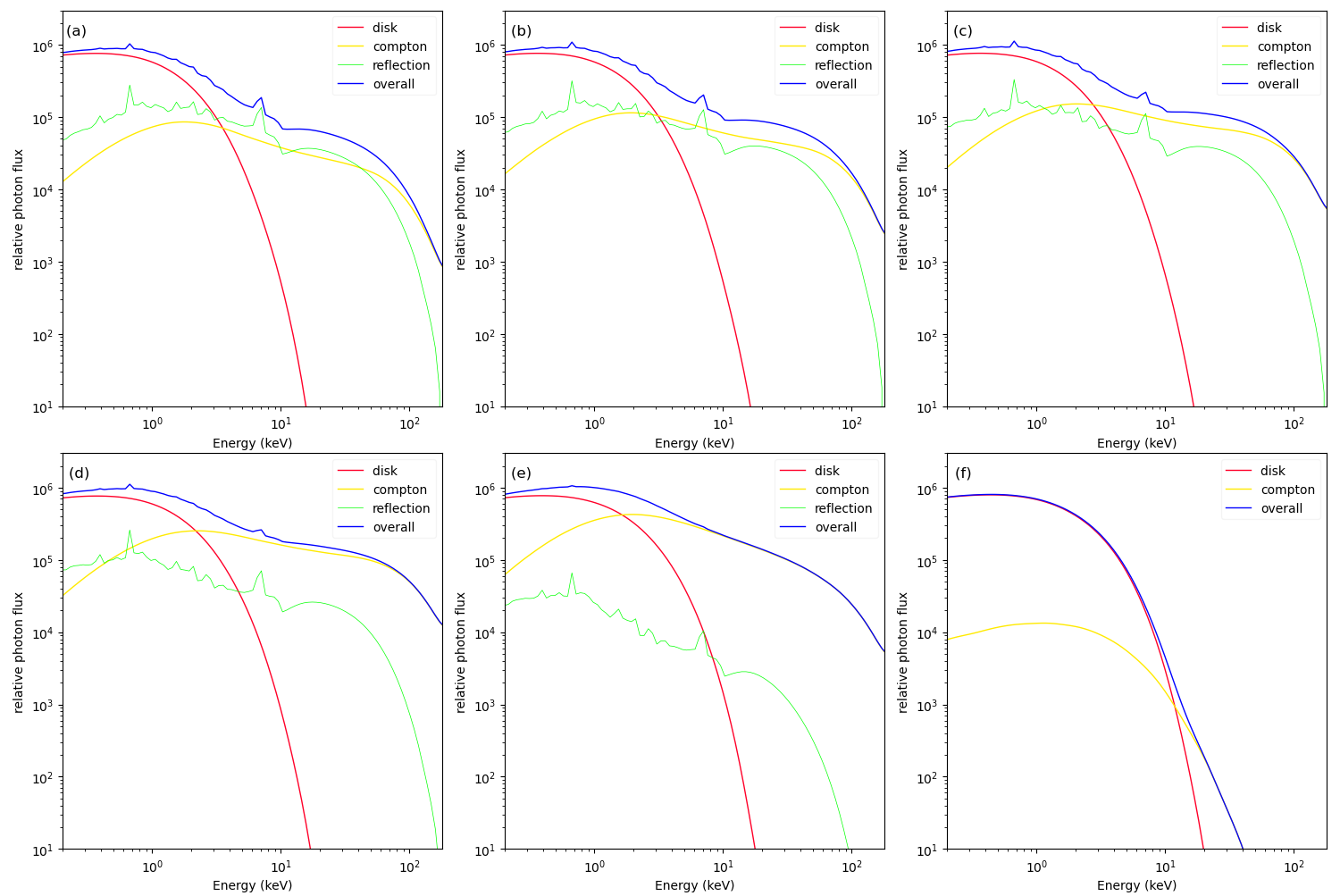}
    \caption{The simulated spectra at inclination angle 45°. The red lines are the thermal component emitted by the accretion disk; the yellow lines are the Compton component; the green lines are the reflection component; the blue lines are the sum of all components, i.e., the observed spectra. Panels (a) to (f) correspond to jet velocity $0 c$, $0.1 c$, $0.2 c$, $0.4 c$, $0.7 c$, $0.99 c$, respectively.}
\end{figure}

\begin{figure}[htbp!]
    \centering
    \includegraphics[width=0.75\linewidth]{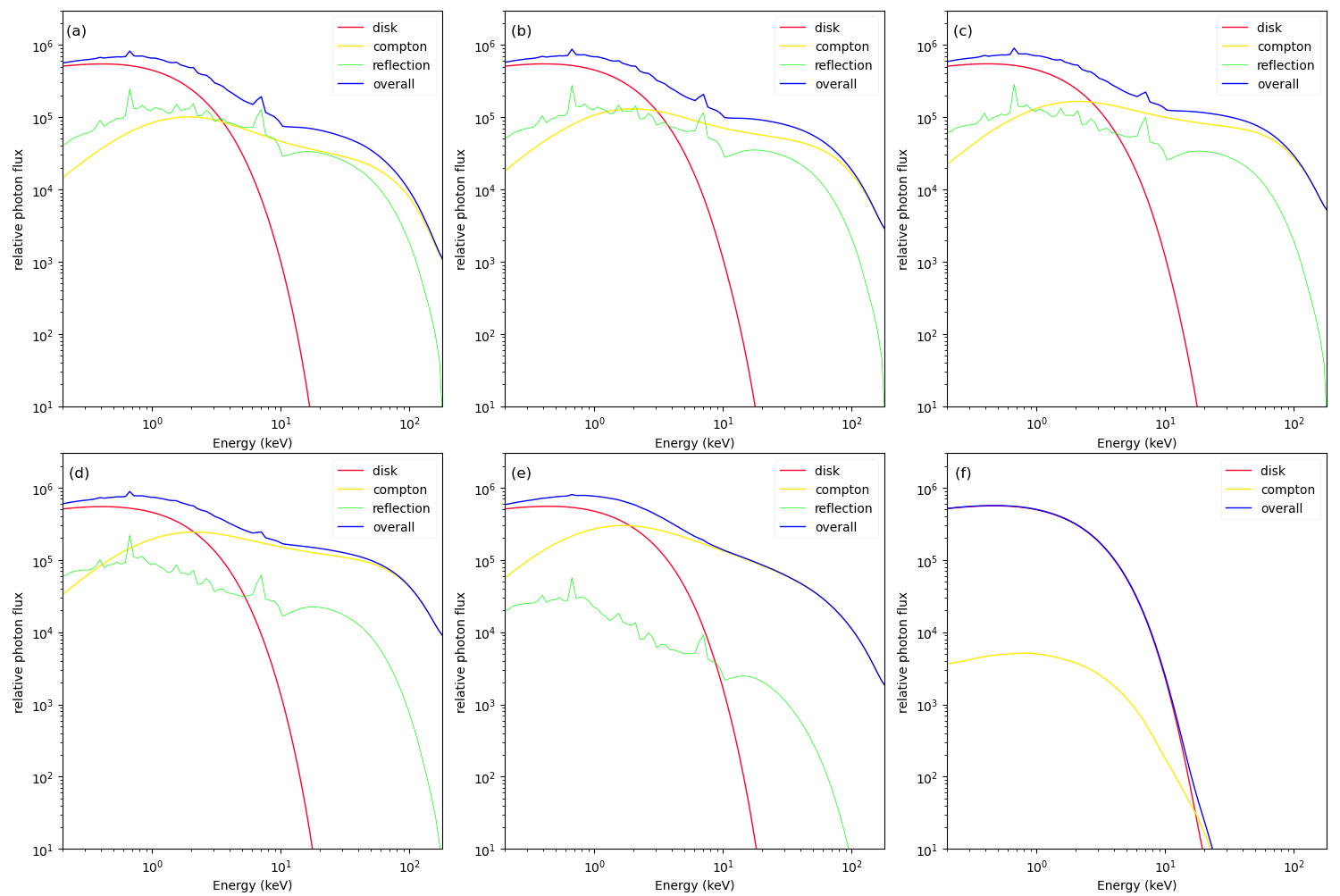}
    \caption{The simulated spectra at inclination angle 55°. The red lines are the thermal component emitted by the accretion disk; the yellow lines are the Compton component; the green lines are the reflection component; the blue lines are the sum of all components, i.e., the observed spectra. Panels (a) to (f) correspond to jet velocity $0 c$, $0.1 c$, $0.2 c$, $0.4 c$, $0.7 c$, $0.99 c$, respectively.}
\end{figure}

\begin{figure}[htbp!]
    \centering
    \includegraphics[width=0.75\linewidth]{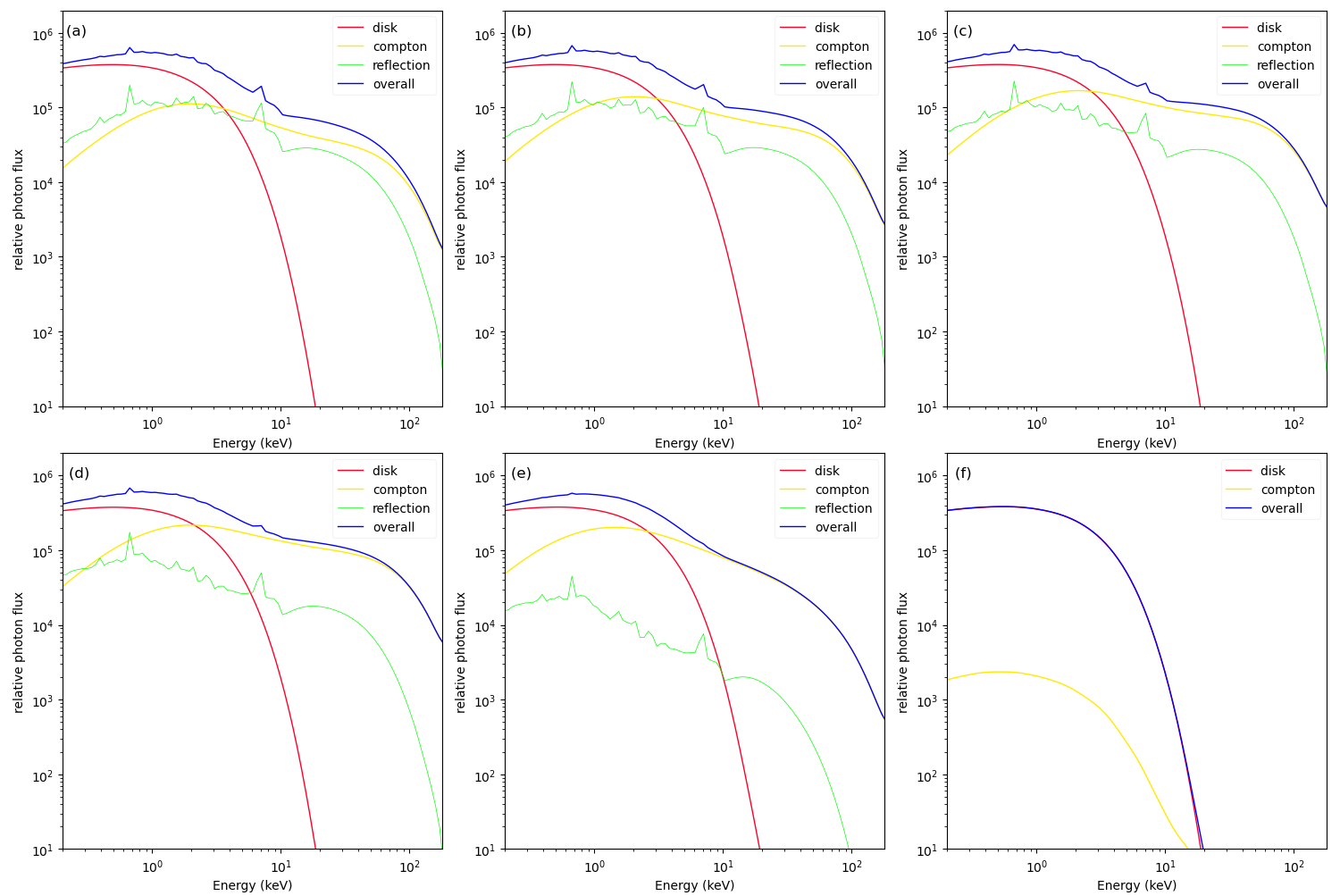}
    \caption{The simulated spectra at inclination angle 65°. The red lines are the thermal component emitted by the accretion disk; the yellow lines are the Compton component; the green lines are the reflection component; the blue lines are the sum of all components, i.e., the observed spectra. Panels (a) to (f) correspond to jet velocity $0 c$, $0.1 c$, $0.2 c$, $0.4 c$, $0.7 c$, $0.99 c$, respectively.}
\end{figure}

\begin{figure}[htbp!]
    \centering
    \includegraphics[width=0.75\linewidth]{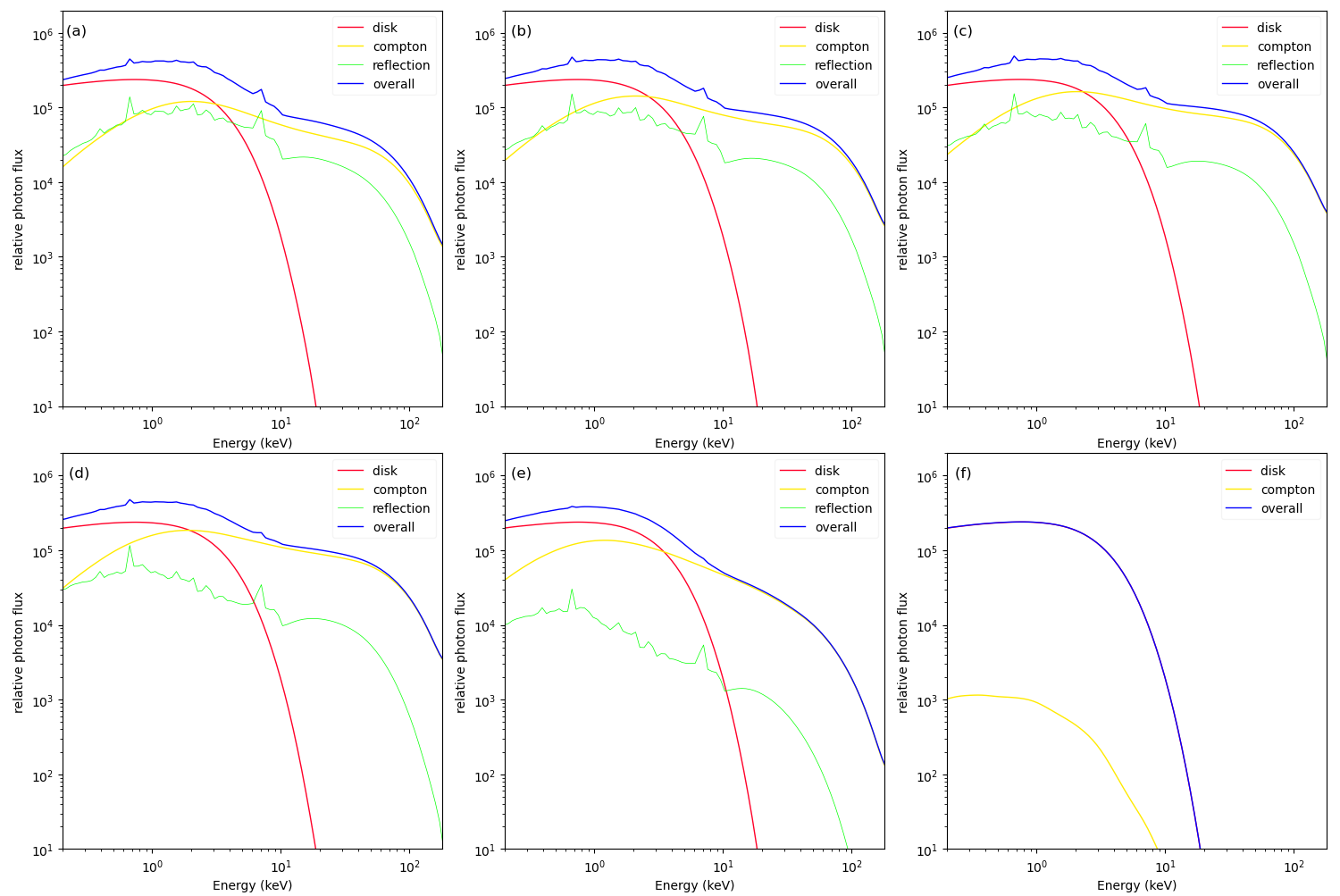}
    \caption{The simulated spectra at inclination angle 75°. The red lines are the thermal component emitted by the accretion disk; the yellow lines are the Compton component; the green lines are the reflection component; the blue lines are the sum of all components, i.e., the observed spectra. Panels (a) to (f) correspond to jet velocity $0 c$, $0.1 c$, $0.2 c$, $0.4 c$, $0.7 c$, $0.99 c$, respectively.}
\end{figure}

\begin{figure}[htbp!]
    \centering
    \includegraphics[width=0.75\linewidth]{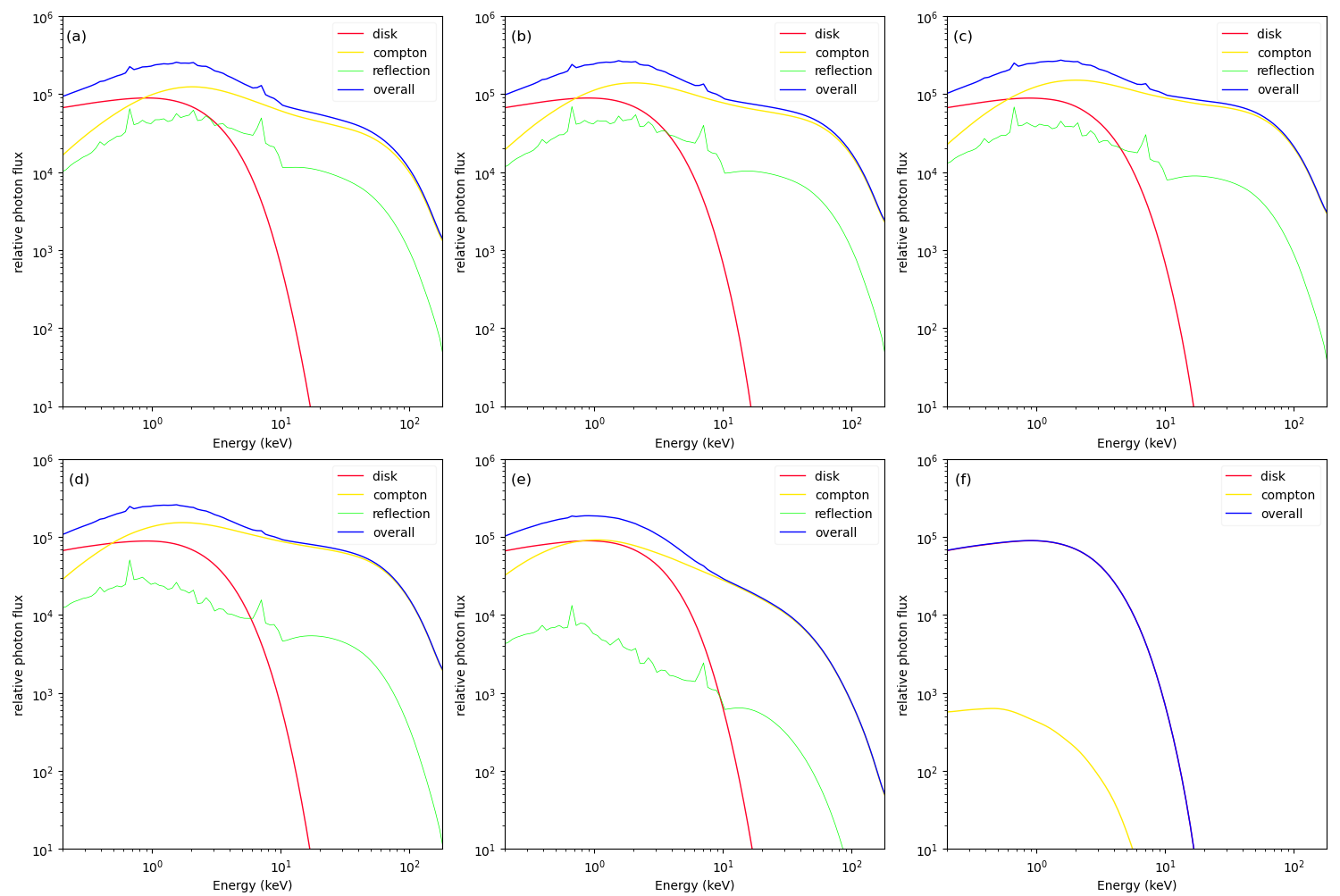}
    \caption{The simulated spectra at inclination angle 85°. The red lines are the thermal component emitted by the accretion disk; the yellow lines are the Compton component; the green lines are the reflection component; the blue lines are the sum of all components, i.e., the observed spectra. Panels (a) to (f) correspond to jet velocity $0 c$, $0.1 c$, $0.2 c$, $0.4 c$, $0.7 c$, $0.99 c$, respectively.}
\end{figure}

\clearpage
\section{the input spectrum of \textbf{xilconv}}
\label{xilconv cut-off}

\begin{figure}[htbp!]
    \centering
    \includegraphics[width=1.0\linewidth]{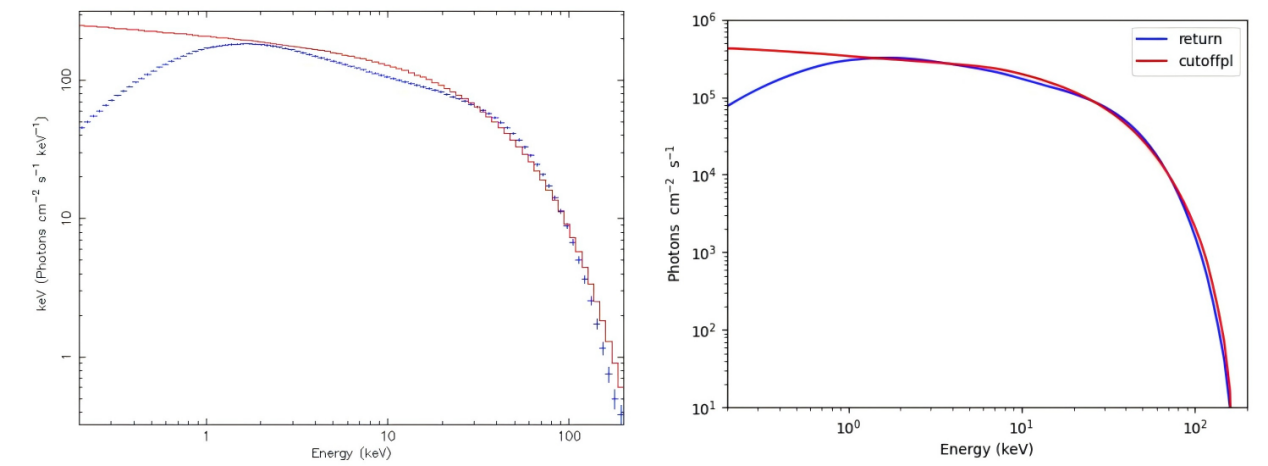}
    \caption{The left panel shows comparisons between a cutoffpl (red) and our input spectrum (blue). The right panel shows the reflection spectra obtained by convolving both the cutoff power law spectrum and the return-to-disk photons spectrum with the xilconv model.}
    \label{input_cutoffpl}
\end{figure}

The XSPEC model \textbf{xilconv} needs an input spectrum that has approximately power-law shape \citep{Garc2022}; however, the typical shape of our input spectrum is a power law with both high and low energy cutoffs. We compare our double-cutoff input spectrum with a standard \textbf{cutoffpl} input spectrum on the calculation of \textbf{xilconv}. As shown in Figure \ref{input_cutoffpl}, the double-cutoff trend of our input spectrum can be catched properly by the calculation result of \textbf{xilconv}. This means that the shape and the intensity of our reflection spectra align with the qualitative analysis of the reflection. Thus, it is not problematic for our simulation to use \textbf{xilconv} in the reflection calculation, considering that it is not a strong reflection situation.

\clearpage
\section{MC simulated spectra with log(Xi)=4.5}\label{appendix A}

In this appendix, MC simulated spectra of log(Xi)=4.5 at inclinations 35°, 45°, 65° and 75° are shown. Spectra of inclinations 25°, 55° and 85° with all other parameters the same are already shown in the main body, corresponding to Figures \ref{fig4}-\ref{fig6}.

\begin{figure}[htbp!]
    \centering
    \includegraphics[width=0.75\linewidth]{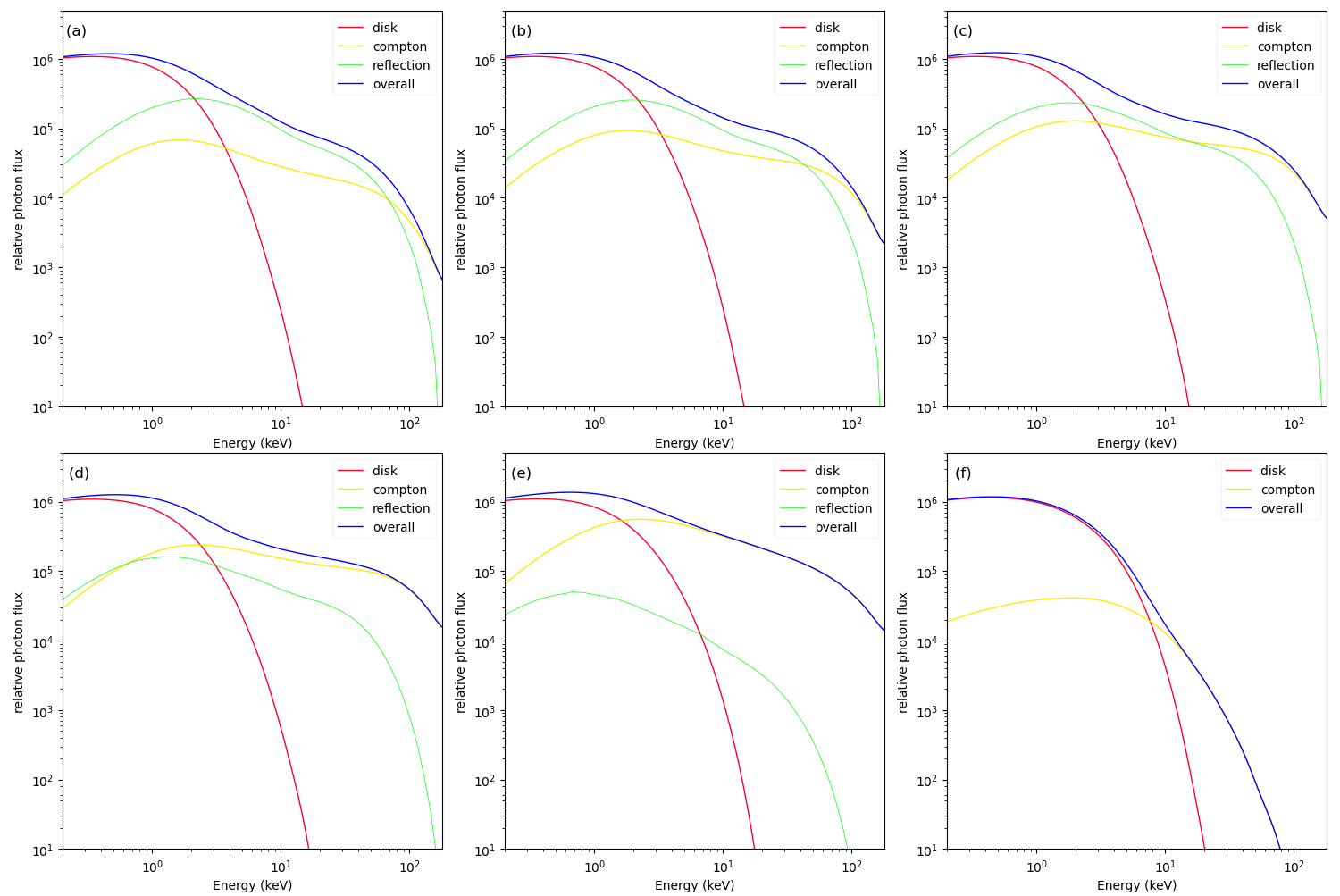}
    \caption{The simulated spectra at inclination angle 35°. The red lines are the thermal component emitted by the accretion disk; the yellow lines are the Compton component; the green lines are the reflection component; the blue lines are the sum of all components, i.e., the observed spectra. Panels (a) to (f) correspond to jet velocity $0 c$, $0.1 c$, $0.2 c$, $0.4 c$, $0.7 c$, $0.99 c$, respectively.}
\end{figure}

\begin{figure}[h!]
    \centering
    \includegraphics[width=0.75\linewidth]{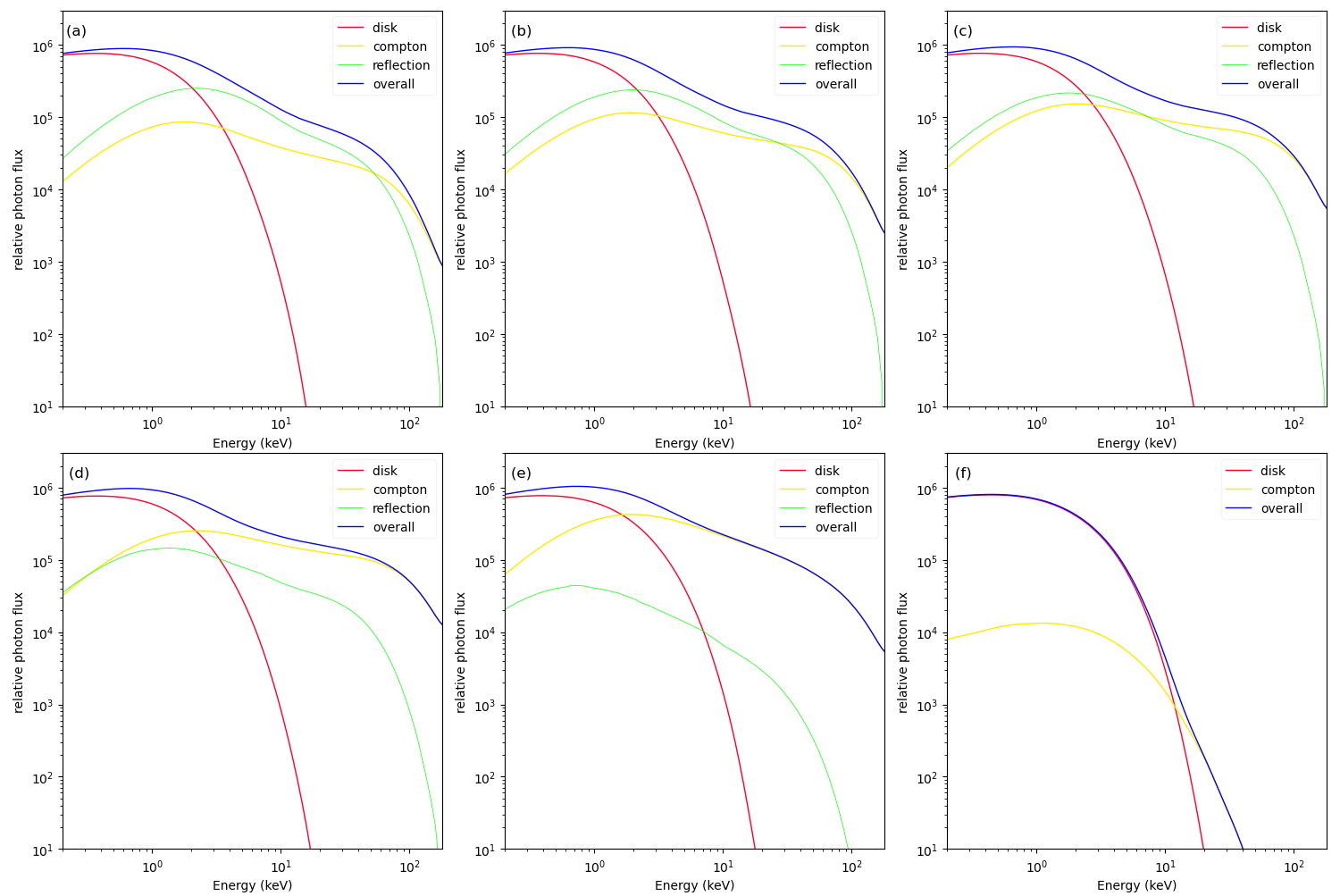}
    \caption{The simulated spectra at inclination angle 45°. The red lines are the thermal component emitted by the accretion disk; the yellow lines are the Compton component; the green lines are the reflection component; the blue lines are the sum of all components, i.e., the observed spectra. Panels (a) to (f) correspond to jet velocity $0 c$, $0.1 c$, $0.2 c$, $0.4 c$, $0.7 c$, $0.99 c$, respectively.}
\end{figure}

\begin{figure}[h!]
    \centering
    \includegraphics[width=0.75\linewidth]{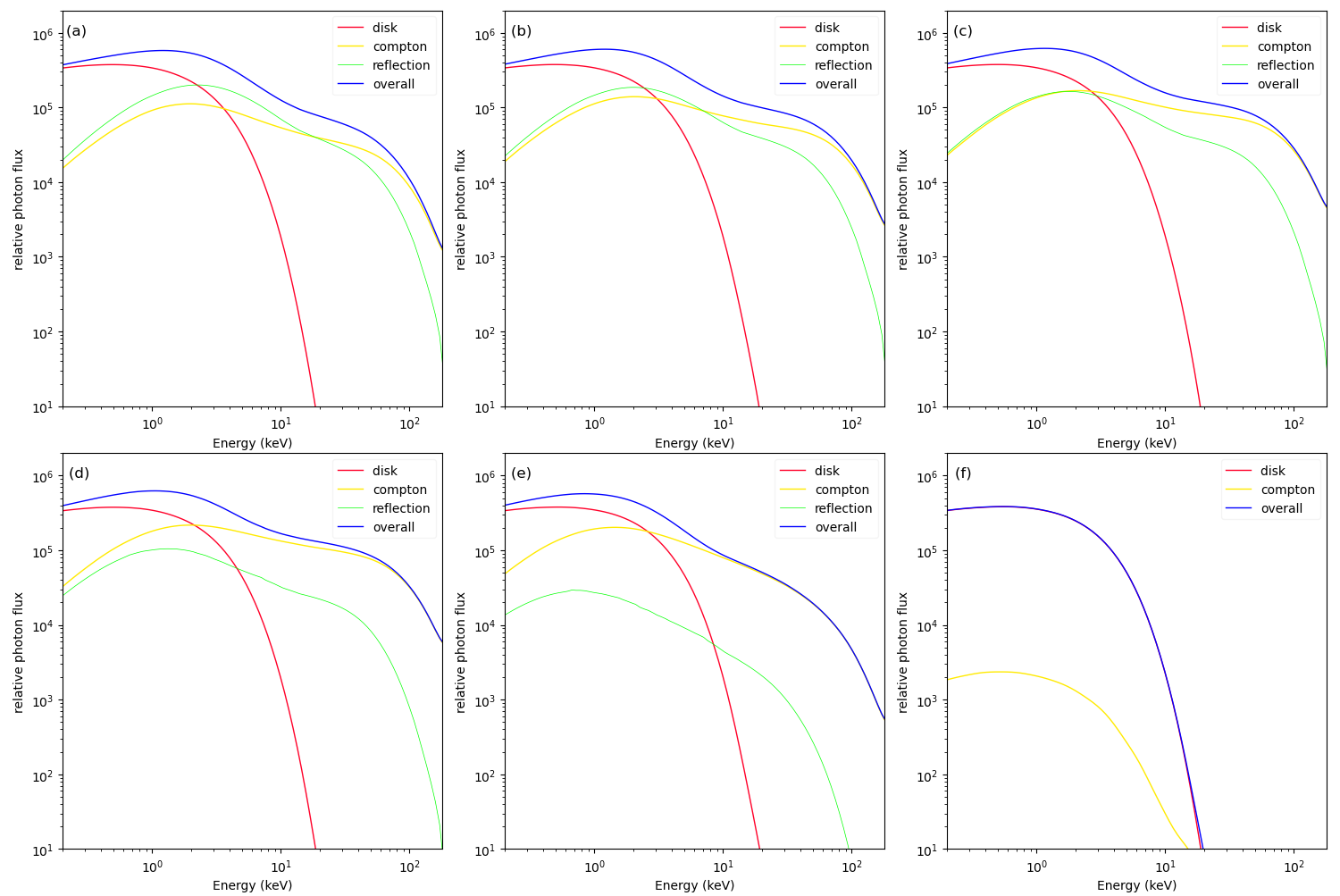}
    \caption{The simulated spectra at inclination angle 65°. The red lines are the thermal component emitted by the accretion disk; the yellow lines are the Compton component; the green lines are the reflection component; the blue lines are the sum of all components, i.e., the observed spectra. Panels (a) to (f) correspond to jet velocity $0 c$, $0.1 c$, $0.2 c$, $0.4 c$, $0.7 c$, $0.99 c$, respectively.}
\end{figure}

\begin{figure}[h!]
    \centering
    \includegraphics[width=0.75\linewidth]{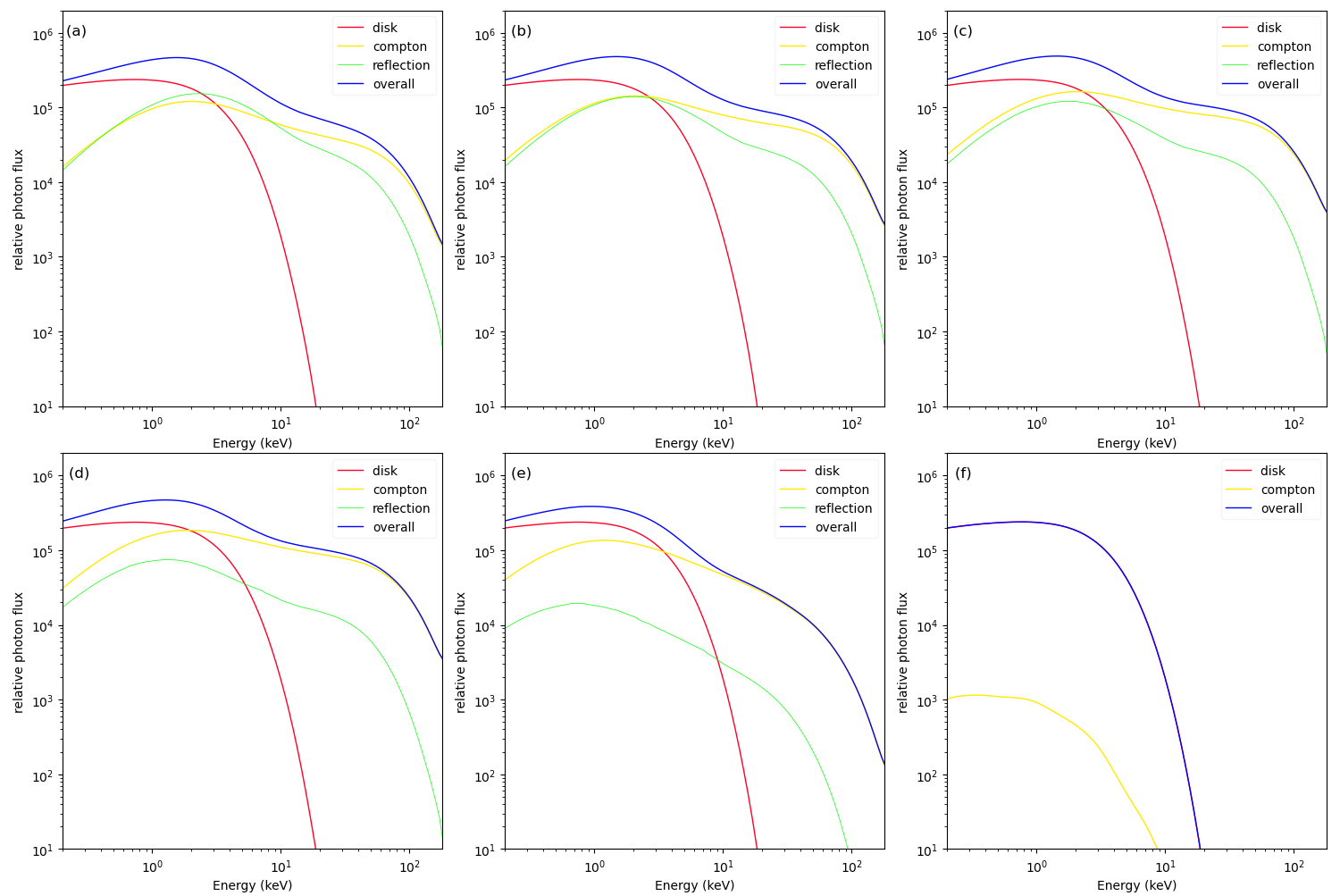}
    \caption{The simulated spectra at inclination angle 75°. The red lines are the thermal component emitted by the accretion disk; the yellow lines are the Compton component; the green lines are the reflection component; the blue lines are the sum of all components, i.e., the observed spectra. Panels (a) to (f) correspond to jet velocity $0 c$, $0.1 c$, $0.2 c$, $0.4 c$, $0.7 c$, $0.99 c$, respectively.}
\end{figure}

\end{document}